\documentclass[twocolumn,pra,nofootinbib,superscriptaddress,amsmath,amssymb,longbibliography]{revtex4-1}
\allowdisplaybreaks[4]
\usepackage{graphicx}
\usepackage{subfigure}
\usepackage{chngcntr}
\usepackage{bm}
\usepackage{soul}
\usepackage[english]{babel}

\usepackage[colorlinks,linkcolor=blue,anchorcolor=blue,citecolor=blue,urlcolor=blue,bookmarksnumbered]{hyperref}
\usepackage{float}

\usepackage{epsfig}
\usepackage{amsmath}
\usepackage{amssymb}
\usepackage{graphicx}

\begin{document}
\linespread{1}\selectfont

\title{Accelerating relaxation through Liouvillian exceptional point}
\author{Yan-Li Zhou}\email{ylzhou@nudt.edu.cn}
\affiliation{Institute for Quantum Science and Technology, College of Science, National University of Defense Technology, Changsha 410073, China}
\affiliation{Hunan Key Laboratory of  Mechanism and technology of Quantum Information, Changsha 410073, China}
\affiliation{Hefei National Laboratory, Hefei 230088, Anhui, China}
\author{Xiao-Die Yu}
\affiliation{College of Science, National University of Defense Technology, Changsha 410073, China}
\author{Chun-Wang Wu}
\affiliation{Institute for Quantum Science and Technology, College of Science, National University of Defense Technology, Changsha 410073, China}
\affiliation{Hunan Key Laboratory of  Mechanism and technology of Quantum Information, Changsha 410073, China}
\affiliation{Hefei National Laboratory, Hefei 230088, Anhui, China}
\author{Xie-Qian Li}
\affiliation{Institute for Quantum Science and Technology, College of Science, National University of Defense Technology, Changsha 410073, China}
\author{Jie Zhang}
\affiliation{Institute for Quantum Science and Technology, College of Science, National University of Defense Technology, Changsha 410073, China}
\affiliation{Hunan Key Laboratory of  Mechanism and technology of Quantum Information, Changsha 410073, China}
\affiliation{Hefei National Laboratory, Hefei 230088, Anhui, China}
\author{Weibin Li}\email{weibin.li@nottingham.ac.uk}
\affiliation{School of Physics and Astronomy, and Centre for the Mathematics and Theoretical Physics of Quantum Non-equilibrium Systems, University of Nottingham, Nottingham NG7 2RD, United Kingdom}
\author{Ping-Xing Chen}\email{pxchen@nudt.edu.cn}
\affiliation{Institute for Quantum Science and Technology, College of Science, National University of Defense Technology, Changsha 410073, China}
\affiliation{Hunan Key Laboratory of  Mechanism and technology of Quantum Information, Changsha 410073, China}
\affiliation{Hefei National Laboratory, Hefei 230088, Anhui, China}

\begin{abstract}
We investigate speeding up of relaxation of Markovian open quantum systems with the Liouvillian exceptional point (LEP), where the slowest decay mode degenerate with a faster decay mode. The degeneracy significantly increases the gap of the Liouvillian operator, which determines the timescale of such systems in converging to stationarity, and hence accelerates the relaxation process. We explore an experimentally relevant three level atomic system, whose eigenmatrices and eigenspectra are obtained completely analytically. This allows us to gain insights in the LEP and examine respective dynamics with details. We illustrate that the gap can be further widened through Floquet engineering, which further accelerates the relaxation process. Finally, we extend this approach to analyze laser cooling of trapped ions, where vibrations (phonons) couple to the electronic states. An optimal cooling condition is obtained analytically, which agrees with both existing experiments and numerical simulations. Our study provides analytical insights in understanding LEP, as well as in controlling and optimizing dissipative dynamics of atoms and trapped ions.
\end{abstract}

\maketitle

\section{Introduction}

Open quantum systems coupled to environments will relax toward a stationary state. The relaxation processes have rich properties from both dynamic and thermodynamic perspectives. Often an important question is to control the relaxation time \cite{Carollo2021, Kochsiek2022,VanVu2021,Lapolla2020}, for instance, on a timescale as short as possible [see Fig. 1(a)]. This problem is of great relevance to cases where one is concerned with properties of stationary states, such as ground state laser cooling \cite{Morigi2000, Feng2020, Double-EIT, Zhang2021b,Zhang2021}, or aims to generate quantum states for quantum applications \cite{Honer2011a,Tresp2016,Paris-Mandoki2017,Stiesdal2020}.

Starting from an arbitrary initial state, the relaxation timescale is largely characterized by the slowest decay mode of the  Liouvillian operator. The gap is defined as modulus of the real part of its eigenvalue $\lambda_1$~\cite{Macieszczak2016a, Znidaric2015, Haga2021}, as depicted in Fig. 1(b). Therefore, relaxation speeding is achieved through increasing the gap. An alternative approach to speed the relaxation is offered by the so-called Mpemba effect \cite{Mpemba_1969, Lu2017, Klich2019, Kumar2020b,Chatterjee2023,Teza2022a}, where an unitary operation on the initial pure state removes its overlap with the slowest decaying mode \cite{Carollo2021, Kochsiek2022}. This transformation can be exactly constructed provided that the initial state is a pure state.

\begin{figure*}[ht]
\includegraphics[scale=0.8]{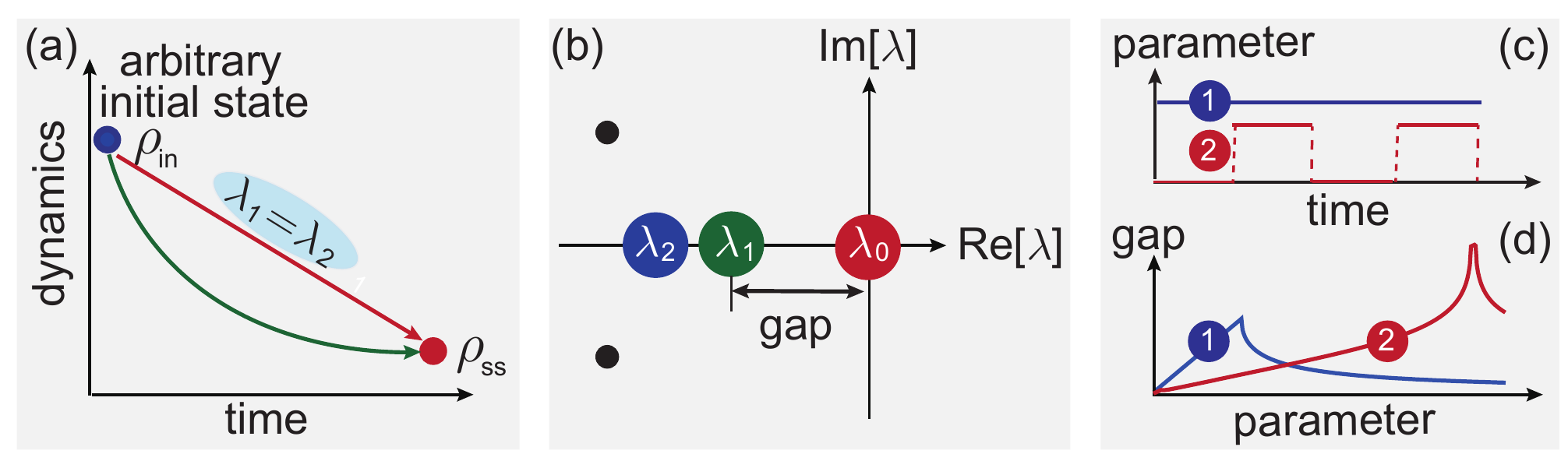}
\caption{(a) For an open quantum system with an arbitrary initial state, the timescale of approaching to the final stationary state $\rho_{ss}$ is related to the slowest decay mode (with eigenvalue $\lambda_1$) of Liouvillian superoperator. By tuning the parameter of the system near the LEPs, where both of the slowest decaying mode and the corresponding eigenvalue are merged with a faster decaying mode and the corresponding eigenvalue (for instance $\lambda_2$), the system dynamics approaches the stationary state in a much faster way. (b) This feature is evident from the Liouvillian spectrum. The stationary state $\rho_{ss}$ is characterized by the largest eigenvalue $\lambda_0 = 0$. The other eigenvalues, characterizing the decay modes, have non-positive real part and always appear as complex conjugates. The Liouvillian spectral gap ($g = -\mathrm{Re}[\lambda_1]$) determines the relaxation timescale, and can reach its maximum value at LEP. (c-d) The gap at LEP can be further increased by the Floquet method (red line). In contrast with the static case (blue line), the gap under time-periodic modulation can be significantly increased, which means that relaxation process will be accelerated by applying the Floquet method.}
	\label{fig:scheme}
\end{figure*}

In this paper, we show that, for an arbitrary initial state, if the slowest decay mode and its corresponding eigenvalue coalesce with a faster decay mode, one can maximize the gap and thus accelerates dynamics to reach stationary states. Our study exploits the novel nature of exceptional points (EPs), which are hallmarks of non-Hermitian systems \cite{Heiss2004,Bergholtz2021,Ashida2020, Minganti2019, Hamazaki2021, Kawabata2019, Miri2019, Wang2021}. EPs are specific points in parameter space, where two or more eigenvalues of a non-Hermitian operator and their corresponding eigenvectors coalesce \cite{Arkhipov2020}. The origin of non-Hermiticity is the coupling between the system and the environment. Liouvillian superoperator, which captures the time evolution of an open quantum system, is non-Hermitian. Therefore, it can exhibit EPs (referred to as Liouvillian EPs, or LEPs)~\cite{Minganti2019, Arkhipov2020}. Properties of LEPs, with diverse unusual effects, have attracted considerable attention currently \cite{Chen2021a, Chen2021c, Kumar2022}, such as dissipative phase transition \cite{Nie2021,Minganti2018, Rubio-Garcia2022}, non-Hermitian skin effect \cite{Longhi2020}, signatures of LEPs in the dynamics \cite{Abbasi2022,Khandelwal2021,Zhang2022} and so on.

Here, we focus on analytical understanding and applications of LEPs in quantum control \cite{Kumar2021b,Pick2019}, and further explore how to accelerate the relaxation towards to stationarity in Markovian open quantum systems via LEPs. The basic mechanism underpinning our study is based on the fact that, at LEPs both the slowest decay mode and  corresponding eigenvalue coalesce with a faster decay mode and the corresponding eigenvalue. We show that, when the stationary state of the system is unique and independent on system parameters, one can set the parameters at the LEP to speed up the relaxation process significantly. For certain quantum dynamic processes, such as ground state cooling, how to convergent to stationarity as quickly as possible is often a concern in the actual process of quantum manipulation. In addition, we find that relaxation processes can be further accelerated by periodically modulating the dissipation strength, i.e., the Floquet modulation can overcome the gap limit of the static case and realize faster relaxation [see Fig. 1(c-d)]. We apply our approach to analyze ground state laser cooling based on sideband transitions and electromagnetically induced transparency (EIT). Optimal conditions are obtained analytically, which have been demonstrated in recent trapped ion experiments~\cite{Zhang2021, Double-EIT}. Our study reveals the importance of LEPs in practical applications and provides insights in seeking optimal conditions in quantum control of open quantum systems.

This paper is organized as follows. In Sec. II, we introduce the master equation of  the Markovian open quantum systems. A general framework that connects to its dynamics and eigenmatrices of the Liouvillian superoperator is provided. This provides an intuitive picture to understand the relaxation and the gap. In Sec. III, we study the dynamics of dissipative three-level system. Eigenmatrices and eigenvalues of the corresponding Liouvillian superoperator are obtained analytically. Based on the analytical calculation, we reveal that the relaxation towards to stationary state can be accelerated by exploiting  static and Floquet-modulated LEPs. Next, in Sec. IV applications for ground state laser cooling are demonstrated. Two experimentally relevant scenarios, i.e. sideband cooling and EIT cooling, are examined. Optimal cooling conditions are obtained at corresponding LEPs. We conclude in Sec. V.

\section{Liouvillian gap, dynamics and LEP}

We consider an open quantum system evolving under Markovian dynamics, governed by master equation $\dot\rho(t)=\mathcal{L} \rho(t)$, where the generator $\mathcal{ L}$, normally called Liouvillian superoperator, has the form \cite{Lindblad1976a, Gorini1975}
\begin{eqnarray}
\mathcal{L}\rho  = - i [H,\rho] + \sum_\alpha\big(J_\alpha\rho J_\alpha^{\dagger} - \frac{1}{2}\{J_\alpha^{\dagger}J_\alpha,\ \rho\} \big).
\end{eqnarray}
Here, $\rho(t)$ is the state of the system at time $t$, $H$ is the system Hamiltonian, and $J_\alpha$ are quantum jump operators which provide coupling of the system to the environment. Since the Liouvillian $\mathcal{L}$ acts linearly on $\rho(t)$, one can obtain information about the relaxation in terms of its eigenmatrices $R_i$ and the corresponding complex eigenvalues $\lambda_i$ via the relation $\mathcal L R_i = \lambda_i R_i$. Note that, duo to the Hermiticity of $\mathcal L$, if $\lambda_i$ is complex, $\lambda_i^*$ must also be an eigenvalue of $\mathcal L$ \cite{Minganti2019,Minganti2018,Carollo2021,Huybrechts2020}. Therefore, the eigenvalues are symmetrically distributed with respect to the real axis as shown in Fig. \ref{fig:scheme}(b).

The stationary state of the system under consideration is then given by the density matrix $\rho_{ss}$ such that $\mathcal L \rho_{ss} = 0$, i.e., $\rho_{ss}=R_0$, which corresponds to the zero eigenvalue $\lambda_0 =0$ and is independent of the initial state. If the eigenvalues are ordered by decreasing their real parts, it is known that the negative real parts of the eigenvalues \cite{Can2019}, $\mathrm{Re}[\lambda_{i>0}] < 0$, determine the relaxation rates of the system towards the nonequilibrium stationary state, and the corresponding eigenmatrices $R_{i>0}$ are called decay modes \cite{Albert2014,Znidaric2015}. While the imaginary parts describe the oscillatory processes which may take place. We can then write the time dependence of the density operator from an initial state $\rho_{in}$ as
\begin{eqnarray}
\rho(t) = e^{\mathcal L t} \rho_{in} = \rho_{ss} + \sum_{i\geqslant 1} a_i e^{\lambda_i t} R_i,
\end{eqnarray}
where $a_i = \mathrm{Tr}[L_i \rho_{in}]$ are coefficients of the initial state decomposition into the eigenmatrices of $\mathcal L^{\dagger}$ with $\mathcal L^{\dagger} L_i = \lambda_i^* L_i$. Here $R_i$ and $L_i$ are referred as right and left eigenmatrices (eigenmodes), respectively, and can be normalized by  Tr$[L_i R_j] = \delta_{ij}$. The trace preservation of the dynamics implies that
$\mathrm{Tr}[\rho(t)]= \mathrm{Tr}[\rho_{ss}] =1 = \mathrm{Tr}[L_0R_0]$, and thus $L_0$ is the identity ($L_0 = I$). It also implies that $\mathrm{Tr}[R_{i\geqslant 1}] =0$, which means other right eigenmatrices do not correspond to quantum states. A particular interesting case is when eigenvalue $\lambda_i$ is real, where  the corresponding eigenmatrix can be diagonalized~\cite{Minganti2018}. We can rewrite it as  superposition of eigenstates from the diagonalization \cite{Minganti2018}
\begin{eqnarray}
    R_i \propto R_i^+ - R_i^-,
\end{eqnarray}
with
\begin{eqnarray}
    R_i^+ = \sum_{n}^{p_n\geq0} p_n^i |\psi_n^i\rangle \langle \psi_n^i |, \nonumber\\
    R_i^- = \sum_{n}^{p_n<0} p_n^i|\psi_n^i\rangle \langle \psi_n^i |,
\end{eqnarray}
where $|\psi_n^i\rangle$ are eigenvectors of $R_i$ with eigenvalues $p_n^i$. With this definition, $R_i^{\pm}$ are arranged to proper density matrices. If $\lambda_i$ is complex, one can define a pair of eigenmatrices $R_i + R_i^{\dagger}$ and $i(R_i - R_i^{\dagger})$, then their corresponding eigenvalues are real (i.e. real and imaginary parts of $\lambda_i$). This allows to diagonlize their new eigenmatrices.

A fundamental role in the system dynamics is played by $\lambda_1 (R_1)$, which possesses the slowest decay rate on the condition $a_1 \neq 0$. Then the Liouvillian gap, defined by \cite{Zanardi2015,Macieszczak2016a}
\begin{eqnarray}
g = |\mathrm{Re}[\lambda_1]|,
\end{eqnarray}
is thus an important quantity determining the timescale of the final relaxation to the stationary state. If the slowest decay mode $\lambda_1 (R_1)$ coalesce with a faster decay mode when we set the system parameters at the so-called LEP, where $\lambda_1(R_1) = \lambda_2(R_2)$, the gap will have extreme values $g_\mathrm{LEP}$. Consequently for long times one has
\begin{eqnarray}
 ||\rho(t) - \rho_{ss} ||\propto e^{g_{\mathrm{LEP}} t},
\end{eqnarray}
where $||A|| = \sqrt{\mathrm{Tr}[AA^{\dagger}]}$ is the Hilbert-Schmidt distance. In such a case, the state would relax at the fastest rate with timescale $1/g_{\mathrm{LEP}}$ for arbitrary initial state. As we will demonstrate latter, this is useful in some quantum applications, where long relaxation timescales become impractical or even harmful to the coherence. Therefore how to quickly approach the steady state becomes necessary in these applications                     .

\section{Analytical LEP theory of dissipative three-level system}

\begin{figure}[ht]
\includegraphics[scale=0.68]{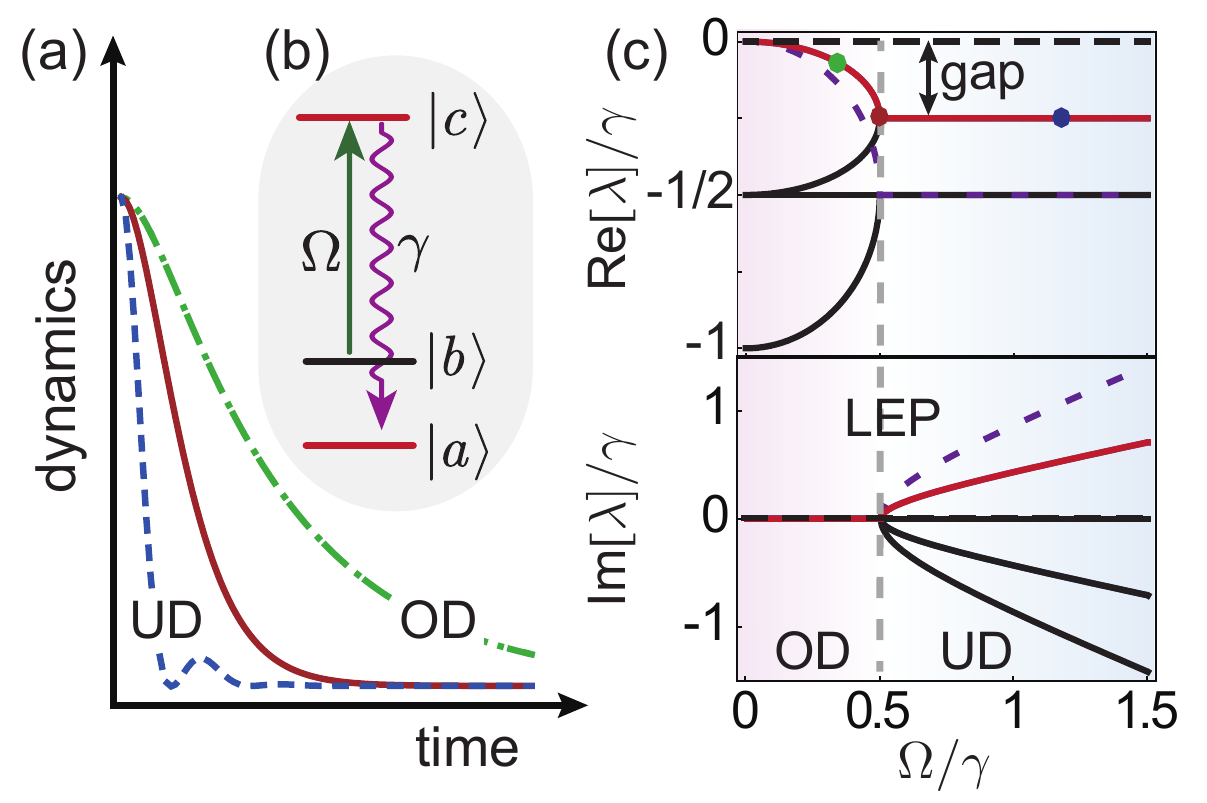}
\caption{(a) Relaxation process. The system decays exponentially to the final stationary state. The numerical integration is performed in the under-damped (UD) regime for $\Omega/\gamma = 0.3$ (dot-dashed green line), in the overdamped (OD) regime with $\Omega/\gamma =1.2 $ (dashed blue line), and in the LEP $\gamma/\Omega = 2$ (solid red line). (b) Schematic of the system, $\gamma$ denotes the emission rate of the $|c\rangle$ level to $|a\rangle$ level, and $\Omega$ denotes the coupling rate from an resonant drive between transition $|b\rangle$-$|c\rangle$. (c) Real and imaginary parts of the Liouvillian spectra of the three-level system shown in (b). The dashed black line corresponds to $\lambda_0=0$. The red solid lines corresponds $\lambda_1$. The Liouvillian gap is denoted by $-\mathrm{Re}[\lambda_1]$, and the LEP is indicated with a vertical dashed line. }
	\label{fig:model_1}
\end{figure}

\subsection{The model}

Consider a simple dissipative three-level system of Fig. ~\ref{fig:model_1}(b). The state $|b\rangle$ resonantly couples to state $|c\rangle$ with Rabi frequency $\Omega$, state $|c\rangle$ decays to state $|a\rangle$ with  decay rate $\gamma$. This process results in state $|b\rangle$ decoupled from the coherent collective evolution and finally decay to the stationary state $|a\rangle$. For the model under consideration, we have the Hamiltonian
\begin{eqnarray}
    H=\frac{\Omega}{2} (|b\rangle \langle c|+ |c\rangle \langle b |)
\end{eqnarray}
and a jump operator $J=\sqrt \gamma|a\rangle \langle c|$. Qualitatively, there will be competition between the reversible coherent coupling between $|b\rangle \leftrightarrow |c\rangle$ at frequency $\Omega$ and the population loss of $|c\rangle$ at a rate $\gamma$. As shown in Fig.~\ref{fig:model_1}(a), if $\Omega \gg\gamma$, Rabi oscillations occur before $|c\rangle$ eventually decays to $|a\rangle$ and dynamics of the system exhibits damped oscillations, which corresponds to under-damped (UD) dynamics. At very long times, the probability in the state $|b\rangle$ tends towards zero since the system ends up in level $|a\rangle$. If, on the other hand, $\Omega \ll\gamma$, we expect an over-damped (OD) evolution for the probability of state $|b\rangle$, which tends exponentially towards zero. The level $|b\rangle$ is irreversibly damped via its coupling with the strongly relaxing level $|c\rangle$, which appears then as an environment for the level $|b\rangle$.

This regime provides a tunable dissipation channel \cite{Honer2011a, Tresp2016,Paris-Mandoki2017, Zhang2022} and recently, it is widely used to simulate parity-time($\mathcal{PT}$)-symmetric Hamiltonians with postselection of the jump results \cite{Naghiloo2019, Wang2021, Chen2021a}. It can also describe the dynamics of the simplest situation of spin-spring system with relaxation processes\cite{Haroche2010}. For instance the damped vacuum Rabi oscillation by the state definition $|a\rangle = |g,0\rangle, |b\rangle = |e,0\rangle$ and $|c\rangle = |g,1\rangle$, where states $|a\rangle$ and $|b\rangle$ are coherently coupled by the Jaynes–Cummings Hamiltonian, while $|b\rangle$ decays towards $|c\rangle$ at the rate $\gamma$. Further more, as we will show below that, this simple model is the core physical principle of phonon ground state cooling \cite{Roos1999, Morigi2000}.

\subsection{The Liouvillian spectra and LEPs}

For this level system, the stationary state of the system is always $|a\rangle$ no matter how the initial state and parameters of the system change. If the goal is to prepare or use this state for related applications, it is unnecessary and even harmful to wait for a long relaxation timescales. In this instance, the most practical construction is to set the optimal parameters to ensure the approaching the stationary state on a timescale which is as short as possible. It can be obtained quantitatively by solving the spectrum of Liouvillian superoperator $\mathcal L$, as shown in Fig. \ref{fig:model_1}(c), and it is $\{0, -(\gamma \pm \kappa )/4, -(\gamma \pm \kappa )/2, -\gamma/2 \}$ with $\kappa=\sqrt{\gamma ^2-4 \Omega ^2}$. Consequently, the spectrum gap is
\begin{eqnarray}
g =\mathrm{Re}[\frac{1}{4} (\gamma - \kappa)],
\end{eqnarray}
and it highlights three regimes for the dynamics as a function of $\Omega/\gamma$ as shown in Fig. \ref{fig:model_1}(c). When $\Omega<\gamma/2$, $\mathcal L$ exhibits a real spectrum, which means that all the excited eigenmodes exponentially decay with time, which corresponds to the OD regime. For $\Omega > \gamma/2$, on the other hand, $\mathcal L$ exhibits a complex spectrum and the system still exhibits Rabi oscillations. For arbitrary initial state, it eventually is damped out with the effectively decay rate which is determined by the gap $g =\gamma/4$. This is UD regime. When $\Omega=\gamma/2$, two pairs of eigenvalues and eigenvectors of Liouvillian coalesce simultaneously (see Appendix A for the exact form of the eigensystem of Liouvillian $\mathcal L$), giving rise to two second-order and a third-order LEPs. It corresponds to a critical damping making the boundary between the OD and the UD regime\cite{Minganti2019, Khandelwal2021}. In particular, at LEP, the gap reaches the maximum value $g_{\mathrm{max}} = \gamma/4 = \Omega/2$, corresponding to that the dynamics at the LEP situation is fastest.

As shown in Fig. \ref{fig:model_1}(c), structures $\{-(\gamma\pm\kappa)/4\}$ and $\{-(\gamma\pm\kappa)/2, -\gamma/2\}$ of the spectrum reflect the two possible relaxation times of the system. Both of them get their minima value when $\kappa =0$, $\lambda_{1(2)}=\lambda_{3(4)}$, $\lambda_5=\lambda_6=\lambda_7=\lambda_8$, giving rise to two second-order  and a third-order LEP. Note that when $\kappa =0$, $R_1 = R_3$, $R_2 = R_4$, $R_5 = R_6 = R_7 \neq R_8$, which means that the corresponding eigenmatrix $R_8$ can not coalesce with $R_{5,6,7}$. Therefore, $\lambda_8$ does not play a role in the $\kappa = 0$ LEP.

\begin{figure}[ht]
\includegraphics[scale=0.7]{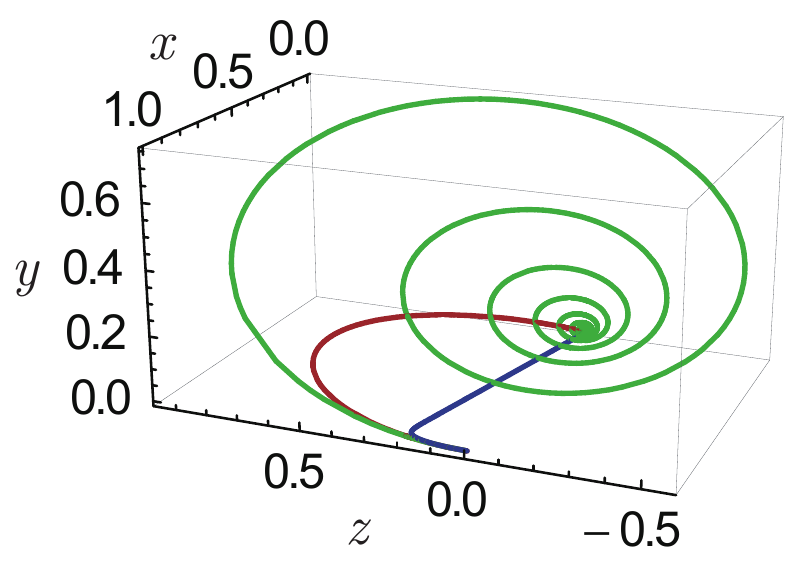}
\caption{The dynamics of $x,y,z$. Green line: oscillatory regime for $\gamma = 0.2 \Omega$, Red line: LEP for $\gamma = 2 \Omega$ and Blue line: overdamped regime for $\gamma = 10 \Omega$.  }
	\label{fig:xyz}
\end{figure}

In order to investigate the physical connotations of the two types of LEPs, we reduce the master equation to the non-zero matrix elements of $\rho$,
\begin{subequations}\label{subequation1}
\begin{align}
\dot{\rho}_{bb} =&\ i \frac{\Omega}{2} (\rho_{bc} - \rho_{cb}),\label{suba}\\
\dot{\rho}_{cc} =&\ -\gamma \rho_{cc} - i \frac{\Omega}{2} (\rho_{bc
} - \rho_{cb}),\label{subb}\\
\dot{\rho}_{bc} =&\ -\gamma/2 \rho_{bc} + i\frac{\Omega}{2} (\rho_{bb} - \rho_{cc}),\label{subc}\\
\dot{\rho}_{aa} = & \gamma \rho_{cc}.\label{subd}
\end{align}
\end{subequations}
The coherent terms couple the populations $\rho_{bb}$ and $\rho_{cc}$ to the coherence $\rho_{bc}$ and $\rho_{cb}$, but have no contribution to the dynamics of coherence $\rho_{bc}+\rho_{cb}$, because $\dot{\rho}_{bc}+\dot{\rho}_{cb} = -\gamma/2(\rho_{bc}+\rho_{cb})$, which is only exponentially damped dynamics with decay rate $\gamma/2$. This means that we can not characterize LEP by observe the dynamics of $(\rho_{bc}+\rho_{cb})$. Meanwhile, as shown in Eq. (\ref{subequation1})(a), the damping term, proportional to $\gamma$, does not affect $\rho_{bb}$. It contributes to the decay of $\rho_{cc}$ and to the corresponding increase of $\rho_{aa}$. The competition between the coherent coupling and the dampling term of two states $\{|b\rangle, |c\rangle \}$ induce the third-order LEP (with the average decay rate $\gamma/2$), witch is the phase transition point of passive $\mathcal{PT}$ Hamiltonian \cite{Wang2021}. And their contributions to $|a\rangle$ give rise the second-order LEP (with average decay rate $\gamma/4$ and half rotating frequency of the third-order LEP).

There are only three independent variables in Eqs. (\ref{subequation1}): $x = \rho_{bb}, y = \rho_{cc}$ and $z = -i(\rho_{bc}-\rho_{cb})$, which describe the dynamics of the subsystem $\{|b\rangle, |c\rangle\}$. With these new notations, we find the dynamics,
\begin{eqnarray}
 \left( \begin{array}{c} \dot{x}\\ \dot{y}\\ \dot{z}\end{array}\right) = \left( \begin{array}{ccc} 0 & 0 & -\Omega/2 \\ 0 &-\gamma & \Omega/2 \\ \Omega & -\Omega & -\gamma/2 \end{array}\right) \left( \begin{array}{c} x \\ y \\ z \end{array} \right).
\end{eqnarray}
Eigenvalues of the $3\times 3$ matrix in Eq. (10) are $ -(\gamma \mp \kappa)/2 \  (=\lambda_{5,6} )$ and $-\gamma/2 \ (=\lambda_7)$. The eigenvalues can be real or complex, leading to the two different regimes qualitatively analysed above and inducing the the third-order LEP at $\kappa=0$. Besides that, we also derive their dynamical evolution analytically
\begin{subequations}\label{subequation2}
\begin{align}
x(t) &= \frac{e^{-\frac{1}{2} \gamma  t}}{\kappa ^2}\Bigg[\left(\gamma ^2-2 \Omega ^2\right) \cosh \left(\frac{\kappa  t}{2}\right)\nonumber\\
&+\gamma  \kappa  \sinh \left(\frac{\kappa  t}{2}\right)-2 \Omega ^2\Bigg],\label{sub1}\\
y(t) &= \frac{e^{-\frac{1}{2} \gamma  t}}{\kappa ^2}4 \Omega ^2  \sinh ^2\left(\frac{\kappa  t}{4}\right),\label{sub2}\\
z(t) &= \frac{e^{-\frac{1}{2} (\gamma +\kappa )t}}{\kappa ^2}\Omega  \left[\gamma  \left(e^{\frac{\kappa t}{2}}-1\right)^2+\kappa  \left(e^{\kappa  t}-1\right)\right].\label{sub3}
\end{align}
\end{subequations}
We show the dynamics of $x(t), y(t), z(t)$ in Fig. 3. When the decay rate is weak ($\gamma<2\Omega$), the evolution is described by damped oscillation with decay rates $\gamma/2$ (subsystem) and $\gamma/4$ (full system), respectively. Obviously, increasing $\gamma$, the evolution approaching to the stationary state will become faster, which corresponds to the quantum anti-Zeno effect \cite{Kofman}. When the decay rate $\gamma>2\Omega$, all the eigenvalues are real and the dynamics exhibits an irreversible damping. In the limit of strong decay, $\gamma\gg 2\Omega$, the relaxation time scale is determined by
\begin{eqnarray}
(\gamma-\kappa)/2 \approx \Omega^2/\gamma \ll \gamma,
\end{eqnarray}
so that the system will experience the metastable process for a long time scale when the system appears stationary, before eventually relaxing to $\rho_{ss} = |a\rangle$. This means that the larger $\gamma$ is, the slower the system relaxes, which is a manifestation of the quantum Zeno \cite{Misra, Fischer2001, Naghiloo2019, Chen2021d, Facchi}. Our results show that, for a dissipative system, quantum Zeno and anti-Zeno effects correspond to the dynamical phenomena with strong and weak dissipation strengths, respectively. The LEP is thus the boundary between the quantum Zeno and anti-Zeno regimes and bridges the two previously independent effects \cite{Li2020}.

As we mentioned before, this dynamics also leads to an effective decay from the state $|b\rangle$ to state $|a\rangle$, with the effective decay rate
\begin{eqnarray}
  \gamma_{b\rightarrow a}  \approx \Omega ^2/\gamma.
\end{eqnarray}
The same result can be found in \cite{Reiter2012a, Zhang2022} by employing perturbation theory and adiabatic elimination of states $|c\rangle$ for a weakly driven between $|b\rangle \leftrightarrow |c\rangle$. The above analysis shows that, our dissipative three-level model can be used to engineer decay processes between state $|b\rangle$ and $|a\rangle$ just by tuning the Rabi frequency $\Omega$.

\subsection{Engineering the relaxation dynamics}
\subsubsection{Control Liouvillian dynamics through the initial state}

\begin{figure}[htb]
\includegraphics[scale=0.55]{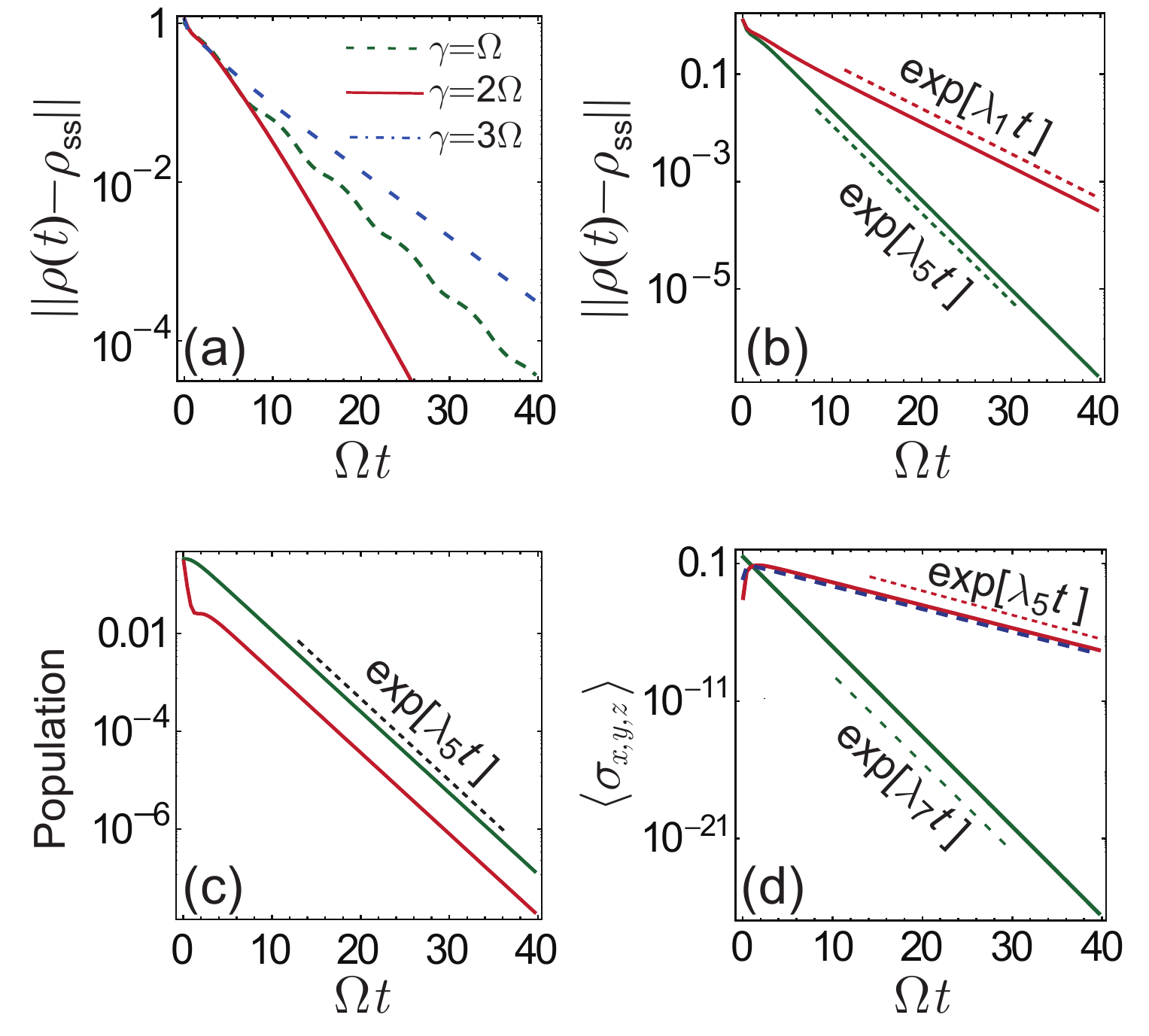}
\caption{(a) Distance between the time-evolved state $\rho(t)$ and the stationary state $\rho_{ss}=|a\rangle$ for an initial random state in the full space with $\gamma = \Omega$ (dashed green line), $\gamma = 2\Omega$ (solid red line), and $\gamma = 3\Omega$ (dot-dashed blue line), respectively. In this original case, the approach to stationary is governed by the eigenvalue $\lambda_1$, and LEP leads to an exponentially faster convergence to the steady state with the rate $g_{\mathrm{LEP}}=\gamma/4 = \Omega/2$. (b) Distance between the time-evolved state $\rho(t)$ and the stationary state $\rho_{ss}$. We compare the case of an initial random state in the full space (red line) with the time evolution ensuing the initial state in the subspace of $\{|b\rangle, |c\rangle\}$ (green line). While in the original case, the approach to stationary is governed by the eigenvalue $\lambda_1$(dashed red line), the special set of initial state leads to an exponentially faster convergence to the steady state with the rate given by $\lambda_5$ (dashed green line). (c) Population dynamics versus evolution time for an initial random state in the full space (green line:$\mathrm{Tr}[\rho(t)|b\rangle\langle b|]$, red line:$\mathrm{Tr}[\rho(t)|c\rangle\langle c|]$), and the time scale is governed by the eigenvalue $\lambda_5$ (dashed line). (d) Observable dynamics versus evolution time for an initial random state in the full space (green line: $\langle \sigma_x\rangle =\mathrm{Tr}[\rho(t)\sigma_x]$, red line: $\langle \sigma_y\rangle=\mathrm{Tr}[\rho(t)\sigma_y]$, dashed blue line: $\langle \sigma_z\rangle=\mathrm{Tr}[\rho(t)\sigma_z]$, and the time scale are different: for $\sigma_x$ it is governed by the eigenvalue $\lambda_7$ (dashed green line) and for $\sigma_{y,z}$ they are governed by the eigenvalue $\lambda_5$ (dashed red line). All the $y$ axises are in logarithmic scale and the parameters for (b-d) are $ \gamma/\Omega = 3$. We have to mention here that the real dynamics and exponential decay function do not coincide at short times. This is because at short time, the decay rate is determined by all decay modes while at long time it decays with time exponentially.}
	\label{fig:decay_time}
\end{figure}

As we discussed in last subsection, there exist two timescales of the relaxation process depending on the space spanned by the initial state. If the initial state is an arbitrary state in space $\{|a\rangle, \{|b\rangle,|c\rangle\}$, the relaxation timescale approaching to the stationary state $|a\rangle$ is determined by $\lambda_1 =- (\gamma - \kappa)/4$, and the fastest dynamical relaxation happens at LEP ($\gamma=2\Omega$) (see Fig. \ref{fig:decay_time}(a)). On the other hand, if the initial state $\rho_{in} \subseteq{\{|b\rangle,|c\rangle\}}$, as shown in Fig. \ref{fig:decay_time}(b), the relaxation timescale is determined by $\lambda_5 =- (\gamma - \kappa)/2$. This means that we can speed up relaxation in the convergence to stationarity by engineering the initial state, which is, the so-called Mpemba effect \cite{Mpemba_1969,Carollo2021,Kochsiek2022}. We can understand this by looking the coefficients $a_i$ of the initial state decomposition into the left eigenmatrices $L_i$. It can be shown that the coefficients of subspace ${\{|b\rangle,|c\rangle\}}$ decomposition into $L_{1\sim4}$ are all vanished, i.e., $a_{1\sim4}=\mathrm{Tr}[L_{1\sim4}\rho_{in}]=0$ (see Appendix A for further
details). In this case
\begin{eqnarray}
\rho(t) = \rho_{ss}+ \sum_{i=5}^8 a_i e^{\lambda_i t}R_i,
\end{eqnarray}
therefore the asymptotic decay rate is $-\mathrm{Re}[\lambda_5] = (\gamma - \kappa )/2$, which can get $g_\mathrm{LEP} = \gamma /2$. In Fig. \ref{fig:decay_time}(b), we compare the timescales for different initial states. It presents that if the initial state is in the full space, the approach to the stationary state is governed by the eigenvalue $\lambda_1$ (red dashed line), while the initial state in the subspace leads to an exponentially faster relaxation to the stationary state with the rate given by $\lambda_5= -(\gamma - \kappa )/2$ (green dashed line).

In addition to the dependence of the relaxation rate on the initial state, we also find that the observable vales have significant effects on the relaxation rate (see Fig. \ref{fig:decay_time}(c,d)). For instance, because $\mathrm{Tr}[|i\rangle\langle i|R_{1\sim 4}] = 0 (i=a,b,c)$, the dynamics of the state populations approaching to stationarity is governed by the eigenvalue $\lambda_5$(Fig. \ref{fig:decay_time}(c)). Moreover, the eigenmatrices $R_{5,6}$ describe the decay of $\sigma_{y,z}$ in the subspace $\{|b\rangle,|c\rangle\}$ with rates $\mathrm{Re}[\lambda_{5,6}]= (\gamma \mp \kappa )/2$, while $R_8 (\lambda_8)$ is associated with $\sigma_x$ (see Appendix A for further details). Whereas there occurs only damped dynamics in the subspace spanned by operators $\sigma_x$, the oscillatory evolution at frequency $|\mathrm{Im}[\lambda_5]|$ in the subspace spanned by vectors $\sigma_{y,z}$  allows to identify the third-order LEP \cite{Minganti2019,Minganti2018,Chen2021c}. Considering that the final state $\rho_{ss} = |a\rangle$ is independent on the parameters of the system, we can speed up the relaxation process by combining the acceleration effect of LEP and the initial state generation.

\subsubsection{Tuning the Liouvillian gap through Floquet modulation}

\begin{figure}[htp]
\includegraphics[scale=0.5]{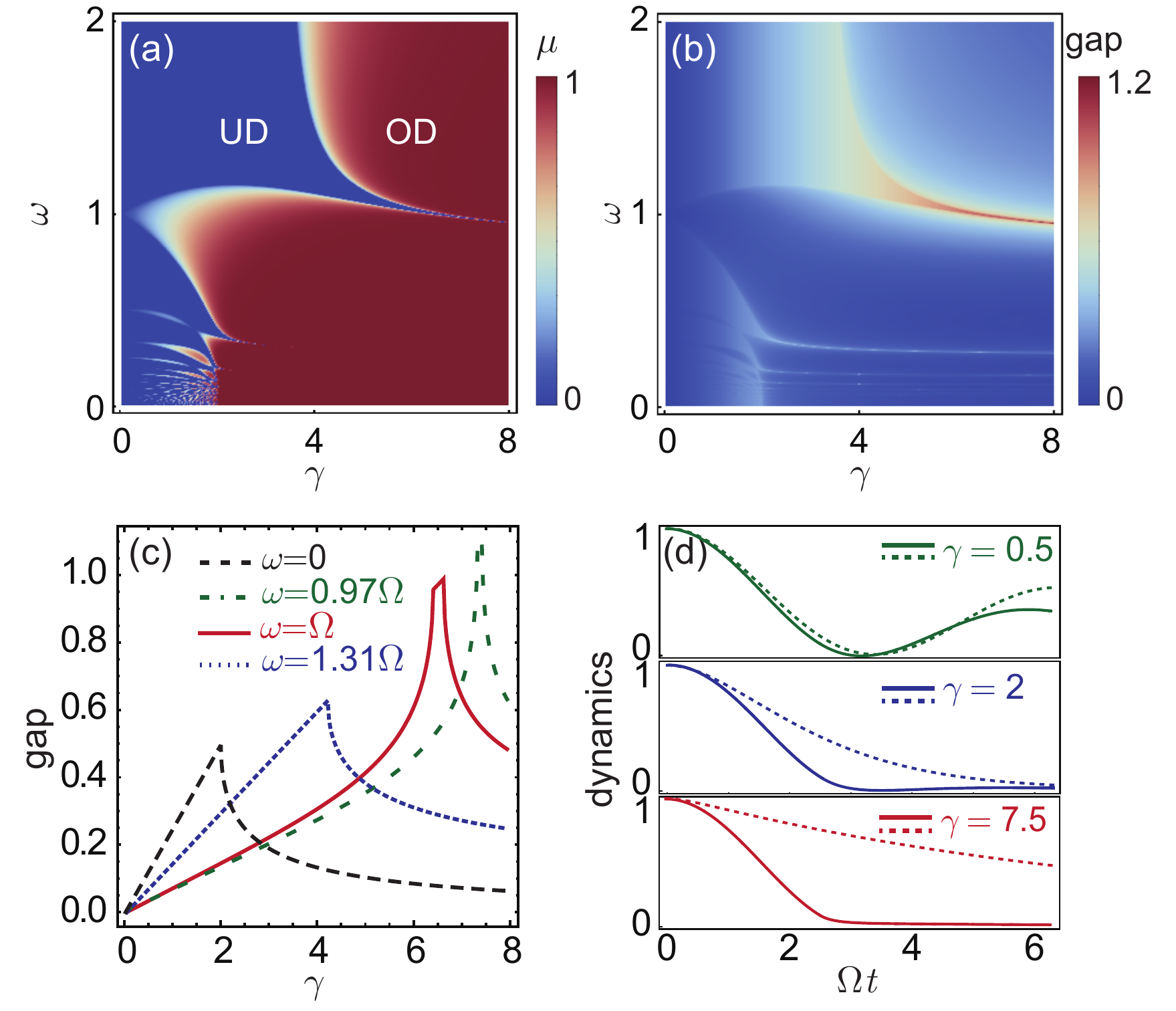}
\caption{(a) Phase diagram in the $\omega-\gamma$ plane. Color contour shows the dimensionless parameter $\mu$. Here we set $\Omega=1, \tau = T/2.5$. Regions with vanishing $\mu$ correspond to the UD regime, while the colored regions correspond to the OD regime. (b) Liouvillian spectral gap as function of $\omega$ and $\gamma$. (c) Liouvillian spectral gap as a function of $\gamma$ with $\omega=0$ (Black dashed line),  $\omega=0.97$ (green dot-dashed line),  $\omega=1$ (red solid line) and $\omega= 1.31$ (blue dotted line), respectively. (d) The dynamics of the population of $|b\rangle$ with $\omega=0$ (dashed lines) and $\omega= \Omega$ (solid lines). The other parameters are $\omega = 0.5$ (green lines), 2 (blue lines) and 7.5 (red lines). }
	\label{fig:floquet}
\end{figure}

We find that the Liouvillian gap can be further increased under time-periodic (Floquet) dissipation with dissipation rate $\gamma$ given by
\begin{eqnarray}
\gamma(t) = \begin{cases}
    0 & n T\leq t < n T + \tau,  \\
     \gamma &  n T + \tau \leq t < (n+1) T.
\end{cases}
\end{eqnarray}
Here $n \in \mathbb Z$, $T= 2\pi/\omega$ is the period of the Liouvillian i.e., $\mathcal L (t+T) = \mathcal L(t)$ with $\omega$ the modulation frequency, and $\tau$ is the off-duty time interval with no decay in each cycle. The density matrix at any time $t$ is determined by the time-evolution operator $\mathcal P(t)= \mathbb T \mathrm{exp}(\int_0 ^t)\mathcal L(t') dt'$, wihere $\mathbb T$ is the time-ordering operator.

In analogy to the case of a non-Hermitian Hamiltonian system \cite{Chen2021d, Li2019}, we can now formally define a Floquet generator for our case, an effective time-independent generator $\mathcal L_F$ such that $\mathcal P(T) = \mathrm{exp}(\mathcal L_F T)$ \cite{Szczygielski2014, Hartmann2017,Schnell2021, Gunderson2021}. Since the UD-OD transitions of the dynamics are determined by the degeneracies of the eigenvalues $\lambda^P$ of the time-evolution operator $\mathcal P$, we adopt a dimensionless parameter $\mu = (|\lambda^P_+| -|\lambda^P_-|)/(|\lambda^P_+| + |\lambda^P_+|)$ to characterize the transition. Here, $\lambda^P_{\pm}$ denote two eigenvalues with bifurcation structure, and $\mu=0$ marks the two eigenvalues are complex conjugate and the system is in the UD regime, while $\mu>0$ are in the OD regime (see Fig. \ref{fig:floquet}(a)). We can see the Floquet method enriches the phase diagram. In contrast with the static dissipation ($\omega = 0$), where the phase transition and LEPs appears at $\Omega/\gamma= 2$, phase transitions under time-periodic dissipation depends on the modulation frequency $\omega$ and can occur at vanishing small dissipation strength.

Beyond that, we are more interested in the effect of modulation on the energy gap. As shown in Fig. \ref{fig:floquet}(b-c), Floquet method increase the gap and the maximal gap appears at the LEP which is a different point with the static case. In the case of static dissipation ($\omega = 0$), the $g_{max} = g_{\mathrm{LEP}}=\Omega/2$ when $\gamma/\Omega= 2$. The gap under time-periodic dissipation depends on the modulation frequency $\omega$ and can even be significantly increased to bigger than $\Omega$ (see Fig. \ref{fig:floquet}(c)). Fig. \ref{fig:floquet}(d) plots the dynamics of the population of state $|b\rangle$. We compare the two different cases, $\omega=0$ (dashed lines) and $\omega=\Omega$ (the solid lines). These results illustrate that the LEP gap can surpass its static limit through Floquet engineering and thus further accelerates the relaxation.

\section{Applications in ground state cooling of trapped ions}

In the following, we demonstrate the power of this approach with a practical application, i.e., the ground state cooling of trapped ions. Through analytical and  numerical analysis, we will illustrate that optimal cooling conditions in the sideband and EIT approaches can be obtained, which agree with existing experiments. Our LEP gap condition provides a new perspective on optimal cooling conditions and may simulate more studies for a wide range of quantum engineering applications.

\subsection{Sideband cooling}

\begin{figure}[htb]
	\includegraphics[scale=0.35]{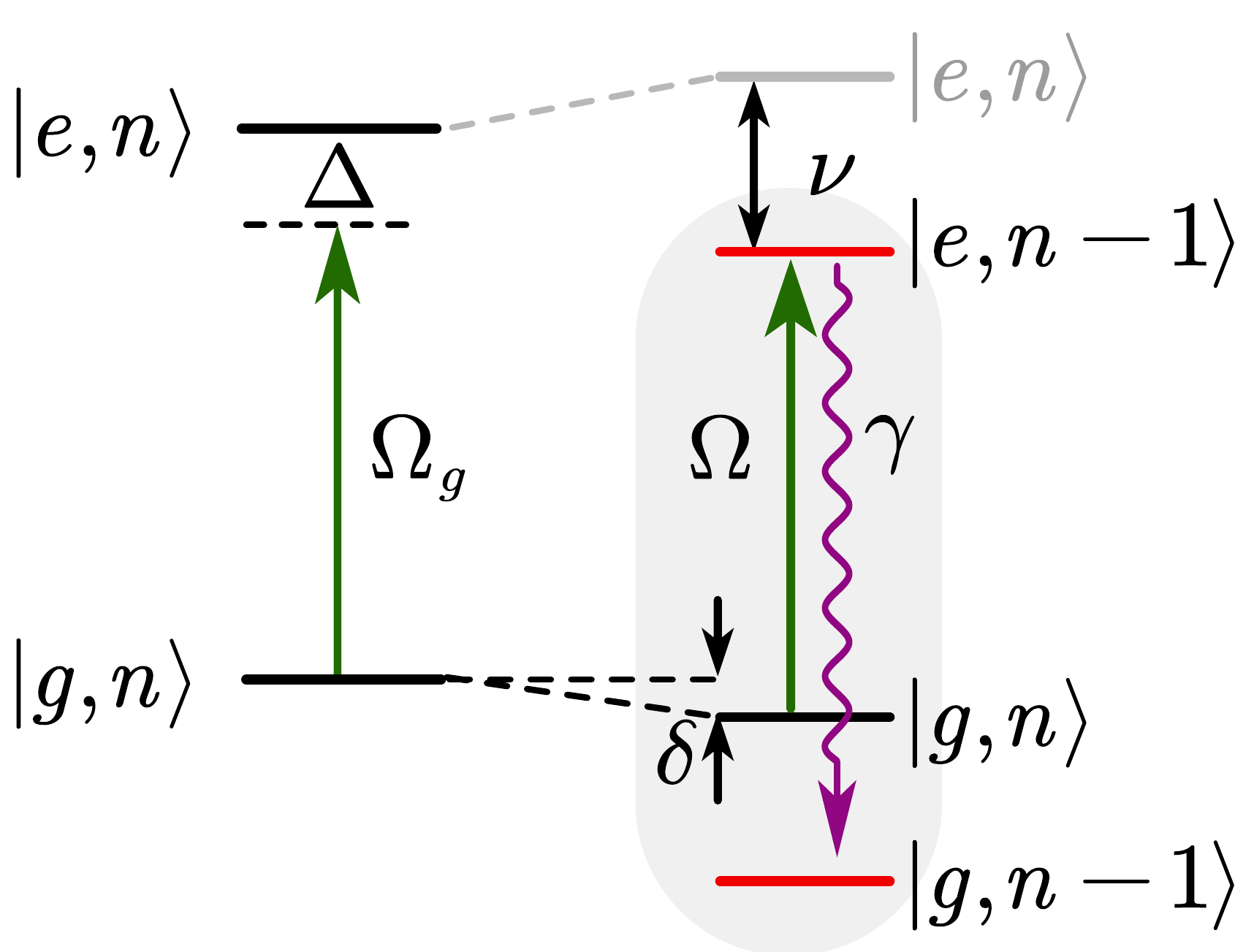}
	\caption{Schematic of the sideband cooling process. The cooling laser with frequency $\omega_l$ drives the transition $|e\rangle\leftrightarrow|g\rangle$ with Rabi frequency $\Omega_g$ and detuning $\Delta=\omega_{eg} - \omega_l$, which leads to the ac Stark shift $\delta$. $\Omega$ is the effective coupling strength between the red sideband transition $|g\rangle |n\rangle \leftrightarrow |e\rangle |n-1\rangle$ with $n$ the phonon number and $\eta$ the Lamb-Dicke parameter.}
	\label{fig:sideband_level}
\end{figure}

As shown in Fig.~\ref{fig:sideband_level}, we consider laser-ion interactions in the Lamb-Dicke limit. Dynamics is governed by Hamiltonian
\begin{eqnarray}
H=\nu a^{\dagger}a + \Delta |e\rangle\langle e|-\frac{1}{2}\Omega_g \sigma^x + \frac{1}{2}\Omega (a^{\dagger} + a)\sigma^y,
\end{eqnarray}
and jump operator $J= \sqrt{\gamma} |g\rangle \langle e|$. $\gamma$ is the linewidth of the state $|e\rangle$, which is coupled to state $|g\rangle$ by a cooling laser field of frequency $\omega_l$, Rabi frequency $\Omega_g$, and detuning $\Delta = \omega_{ge} - \omega_l$, where $\omega_{ge}$ is the frequency of the bare atomic transition $|e\rangle \leftrightarrow |g\rangle$. $\nu$ is the trap frequency and $a (a^{\dagger})$ is annihilation (creation) operator of phonons. $\Omega = \eta \Omega_g$ describes the effective coupling between the phonon and internal state and $\eta$ is the Lamb-Dicke parameter. When $\Delta \simeq \nu$, the red sideband transition is nearly resonant, and the non-resonant transitions, i.e., the carrier transition and blue sideband transition, will induce the ac Stark shift to $|e\rangle (|g\rangle)$ by $\delta (-\delta)$, respectively. This is a good approximation that just considering shift caused by carrier transition, and under this approximation we get $\delta = (\sqrt{\Omega_g^2 + \Delta^2}-\Delta)/2$.

\begin{figure}[htb]
\includegraphics[scale=0.75]{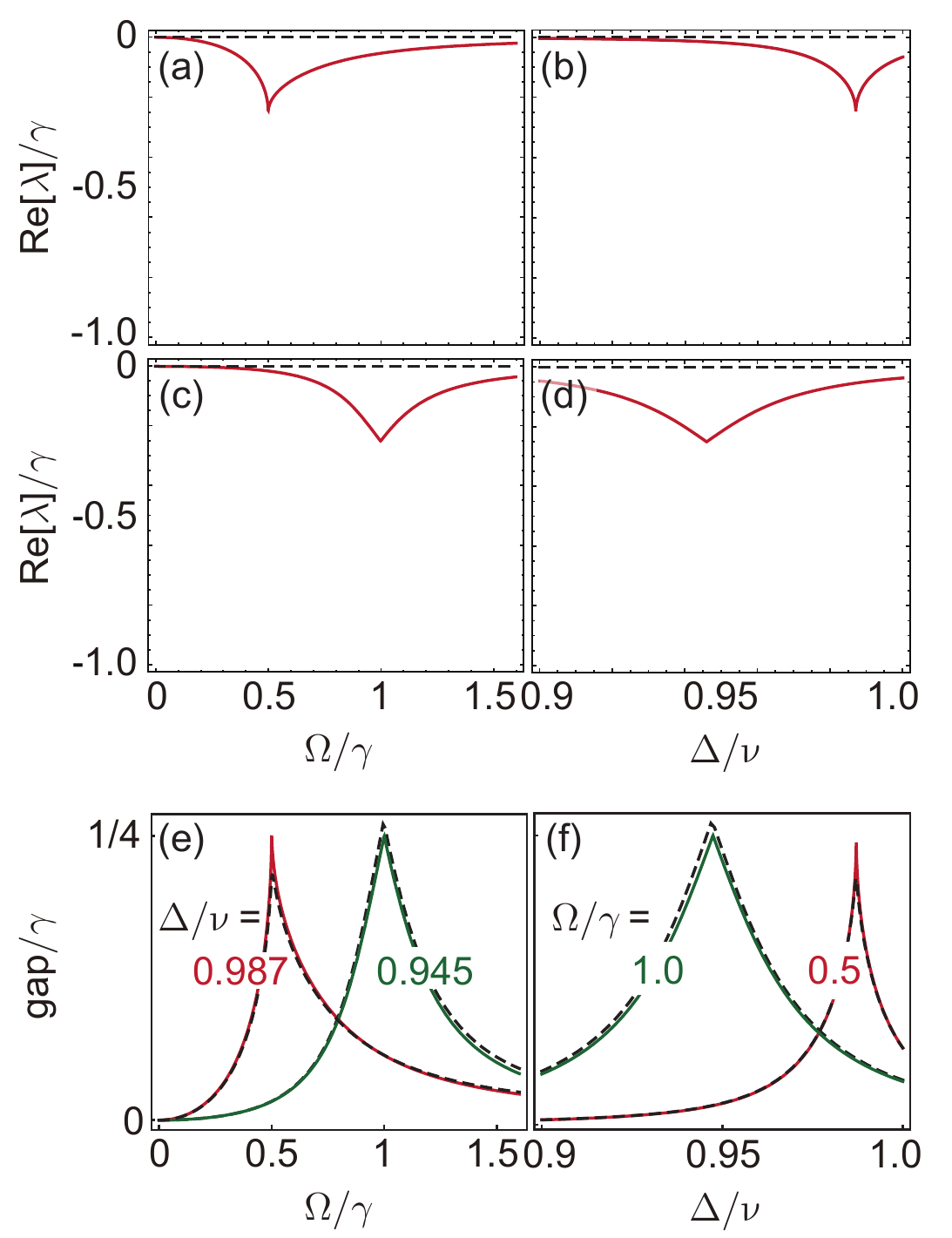}
	\caption{(a-d) Real parts of the Liouvillian spectra of the analytical results. The black dashed lines are $\lambda_0$, the red lines are $\mathrm{Re}[\lambda_1]$ and the green dotted lines are $\mathrm{Re}[\lambda_{i>1}]$. The parameters are (a):
$\Delta = 0.987 \nu$, (b): $\Omega = 0.5 \gamma$, (c): $\Delta = 0.945 \nu$, and (d): $\Omega = \gamma$. (e-f)Gap as functions of $\Omega/\gamma$(e) and $\Delta/\nu$ (f). (e): $\Delta = 0.987\nu$ (red solid line) and $=0.945\nu$ (green solid line), (f): $\Omega = \Omega_g \eta = 0.5\gamma $ (red solid line) and $= \gamma$ (blue solid line), respectively. The solid lines correspond to the analytical results from eq. (\ref{eq:4level}) and the black dashed lines are the results from the full Liouvillian. The other parameters are $\eta = 0.1, \nu =1, \gamma = 0.032$. }
	\label{fig:gap(Delta,Omega)}
\end{figure}

\begin{figure}
\includegraphics[scale=0.55]{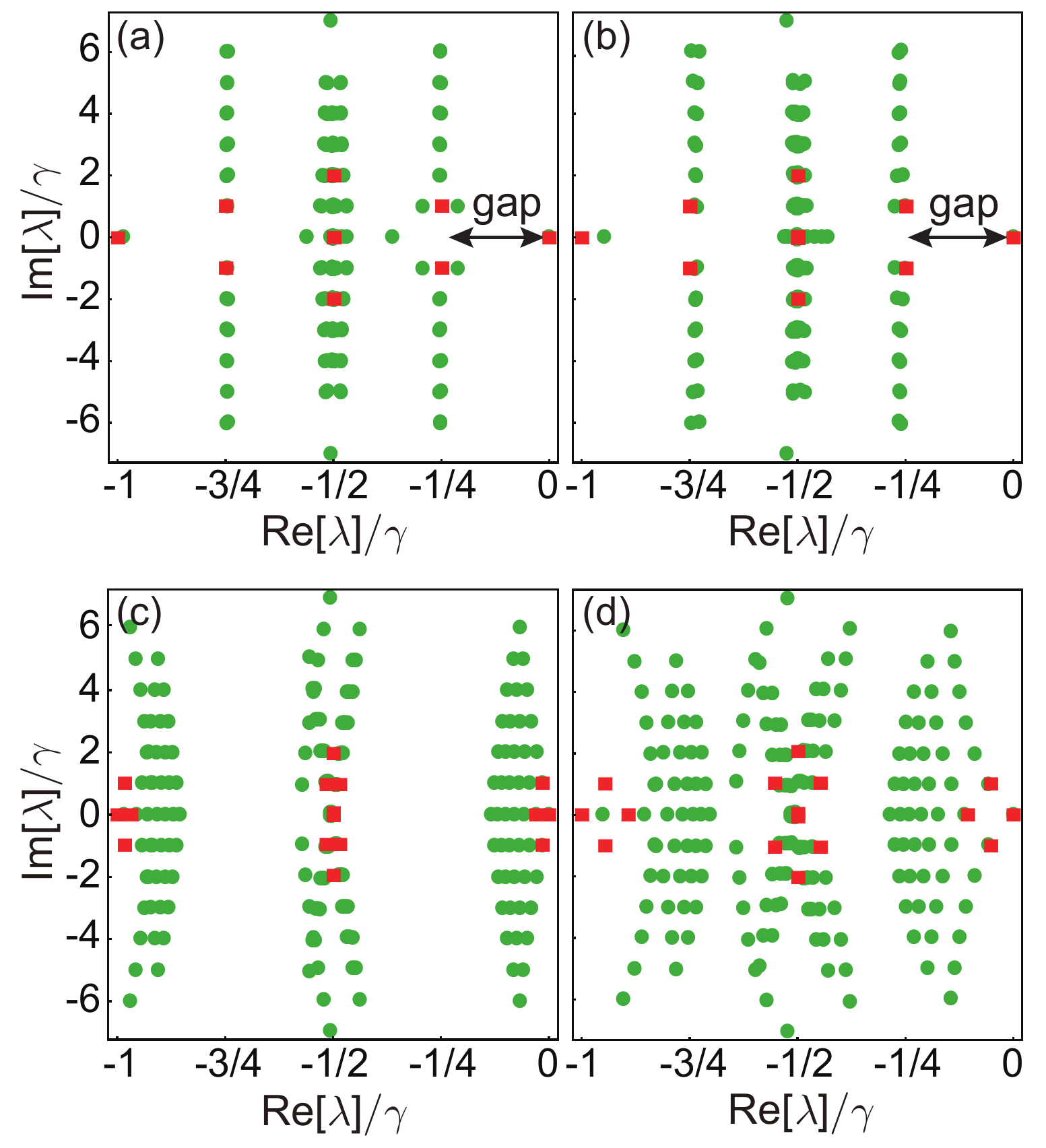}
	\caption{Spectral properties of the Liouvillian. (a-b) in the LEP condition $(\Omega^2 + \Delta = \nu^2)$, and the parameters of $(\Omega/\gamma, \Delta/\nu)$ are (a): $(0.5,0.987)$, (b): $(1,0.945)$, respectively. Fig. (c-d) not in the LEP condition $(\Omega^2 + \Delta \neq \nu^2)$, the parameters  of $(\Omega/\gamma, \Delta/\nu)$ are (c): (1, 0.987), (d): (0.2, 1). The green dots are the results from the full Liouvillian and the red squares are the eigenvalus of the analytical results. The other parameters are the same with Fig. \ref{fig:gap(Delta,Omega)}.}
	\label{fig:spectr}
\end{figure}

\begin{figure}[htb]
\includegraphics[scale=0.5]{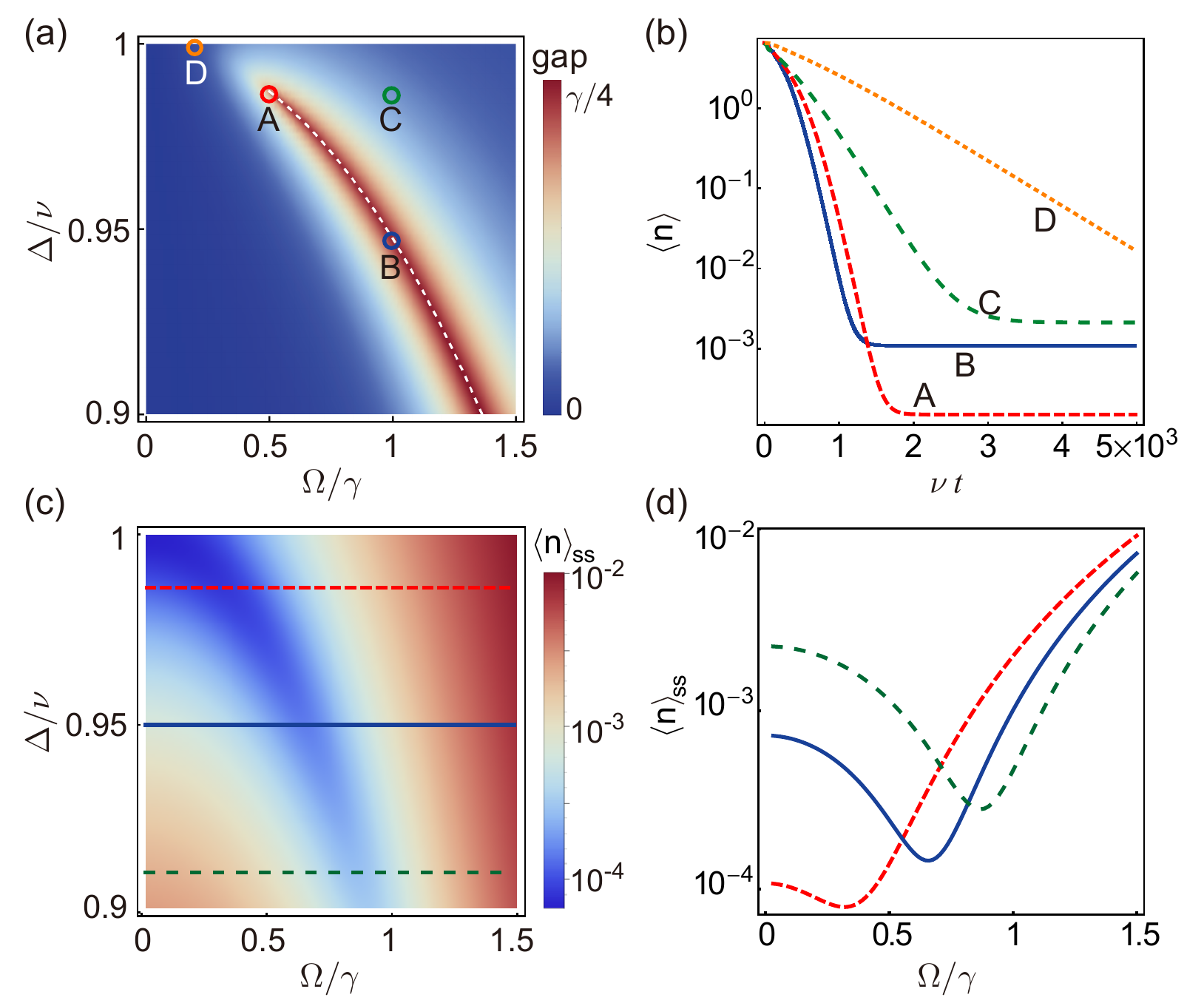}
	\caption{(a) Gap properties of the full Liouvillian. The white dashed line corresponds to $\Omega^2 + \Delta^2 = \nu^2$. The parameters$(\Omega/\gamma, \Delta/\nu)$ are $(0.5,0.987) (A), (1,0.945) (B), (1, 0.987) (C), (0.2, 1) (D),$ and $g_A = g_B > g_C > g_D $. (b) The average phonon number $\langle n \rangle $ as a function of time calculated with full master equation with the parameters of A, B,C,D, respectively. (c) Cooling limit $\langle n \rangle_{ss}$ as functions of detuning $\Delta/\nu$ and effective Rabi frequency of red sideband transition $\Omega/\gamma$. (d) $\langle n \rangle_{ss}$ as a function of effective Rabi frequency of red sideband transition $\Omega/\gamma$ with $\Delta/\nu =0.987$ (red dashed line), $0.95$ (blue solid line) and $0.91$ (green dot-dashed line). The other parameters are the same with Fig. \ref{fig:gap(Delta,Omega)}.}
	\label{fig:gap2d}
\end{figure}

In order to obtain the optimal cooling condition, we reduce the overall dynamics to a low-dimensional subsystem $\{|g\rangle |1\rangle, |g\rangle |0\rangle, |e\rangle |1\rangle, |e\rangle |0\rangle\}$ to obtain analytical results (see Appendix B for details). It is very helpful for us to understand the whole cooling process. Based on the perturbative calculations for this finite systems, we get that $\lambda_{1(3)} =-(\gamma\mp\kappa')/4+ i(\Delta +2 \delta+\nu)/2$ and the gap
\begin{eqnarray}
g = \mathrm{Re}[\frac{1}{4}(\gamma - \kappa')],
\label{eq:4level}
\end{eqnarray}
with $\kappa'= \sqrt{(\gamma -2 i (\Delta +2 \delta-\nu ))^2-4 \Omega ^2}$. In Fig. \ref{fig:gap(Delta,Omega)} (a-d), we plot the analytical results of the spectra. It is obvious that the LEPs can only occur under the condition $\Delta +2 \delta-\nu =0$. With $\delta = (\sqrt{\Omega_g^2 + \Delta^2}-\Delta)/2$, we obtain the condition to generate LEP,
\begin{eqnarray}
\Omega^2 + \Delta^2 =\nu^2.
	\label{eq:condition}
\end{eqnarray}
Under this condition, the eigenvalues $\lambda_{1(3)}$ become $\lambda_{1(3)} = -(\gamma \mp\kappa)/4+i\nu$, whose real parts are the same with three-level dissipative system shown in Fig. \ref{fig:model_1}(b) and the imaginary parts connote rotating. The physical mechanism underlying condition (\ref{eq:condition}) is that the detuning $\Delta$ needs to be adjusted according to the ac Stark shift of the atomic levels to ensure that the red sideband transition is exactly on resonance. Under this premise, the level structure shown in Fig. \ref{fig:sideband_level} can be considered as a simple three-level dissipative system discussed in section III. When $\Omega \geqslant \gamma/2$, $\mathrm {Re} [\lambda_1] = \mathrm {Re} [\lambda_3] = -\gamma/4$, and we get the maximum value $g_{\mathrm{max}} = \gamma/4$. Particularly, when $\Omega=\gamma/2$, $\kappa = \sqrt {\gamma^2 - 4\Omega^2} = 0$, $\lambda_1 = \lambda_3$ and $R_1 = R_3$ (see Appendix B), LEP occurs.

In Fig.\ref{fig:gap(Delta,Omega)} (e-f), we compare the gap given by Eq. (\ref{eq:4level}) (the solied lines) with the numerical results calculated from the full master equation (the dashed lines). Although our analytical results about spectrum and gap are obtained from the subsystem of the sideband cooling, they match very well with the numerical results calculated from the full-system. As shown in Fig. \ref{fig:spectr}, the real parts of the eigenvalues $\lambda_{i\geqslant 1}$, which give the relaxation rates of all the decay modes of the system, mainly can be divided into several characteristic intervals. It approximately could be $\{-\gamma/4,-\gamma/2, -3\gamma/4, -\gamma\}$ under the condition (\ref{eq:condition}) (see Fig. \ref{fig:spectr} (a-b)), otherwise, it becomes to be $\{0, -\gamma/2, -\gamma\}$ when $\Omega/\gamma>1/2$  (see Fig. \ref{fig:spectr} (c)). As shown in Fig. \ref{fig:spectr} (c-d), different with the 3-level dissipation system, it features a so-called metastable regime either $\Omega/\gamma>1/2$ or $\Omega/\gamma<1/2$, which occurs when low lying eigenvalues become separated from the rest of the spectrum \cite{Macieszczak2016a} . The imaginary parts of the eigenvalues, which give the rotating rates of the decay modes approximately equal to $\mathrm{Im}[\lambda_{i\geqslant 1}] \approx (\Delta +2 \delta+n\nu)/2 $ are mainly determined by phonon energy.

In Fig. \ref{fig:gap2d}(a), we plot the gap $g$ as functions of $\Omega/\gamma$ and $\Delta/\nu$ by using the full master equation. The point A corresponds to the LEP, and the dashed white line is the condition of $g = g_{max}$ that combines red sideband transition resonant condition in Eq. (\ref{eq:condition}) as we discussed before. The numerical results and the analytical results are matched very well. Fig. \ref{fig:gap2d}(b) shows the the dynamics of the full system for some set of parameters (the points A, B, C, D in Fig. \ref{fig:gap2d}(a)). It indicates that the gap $g$ provides a good description of the cooling time. And at the LEP, the system not only reaches stationary state at a significantly faster pace, but also obtains a lower phonon number (see Fig. \ref{fig:gap2d}(c-d)). Therefore, we believe that the best cooling effect can be obtained by the system parameters at the LEP.

\subsection{EIT cooling}

\begin{figure}[htb]
\includegraphics[scale=0.23]{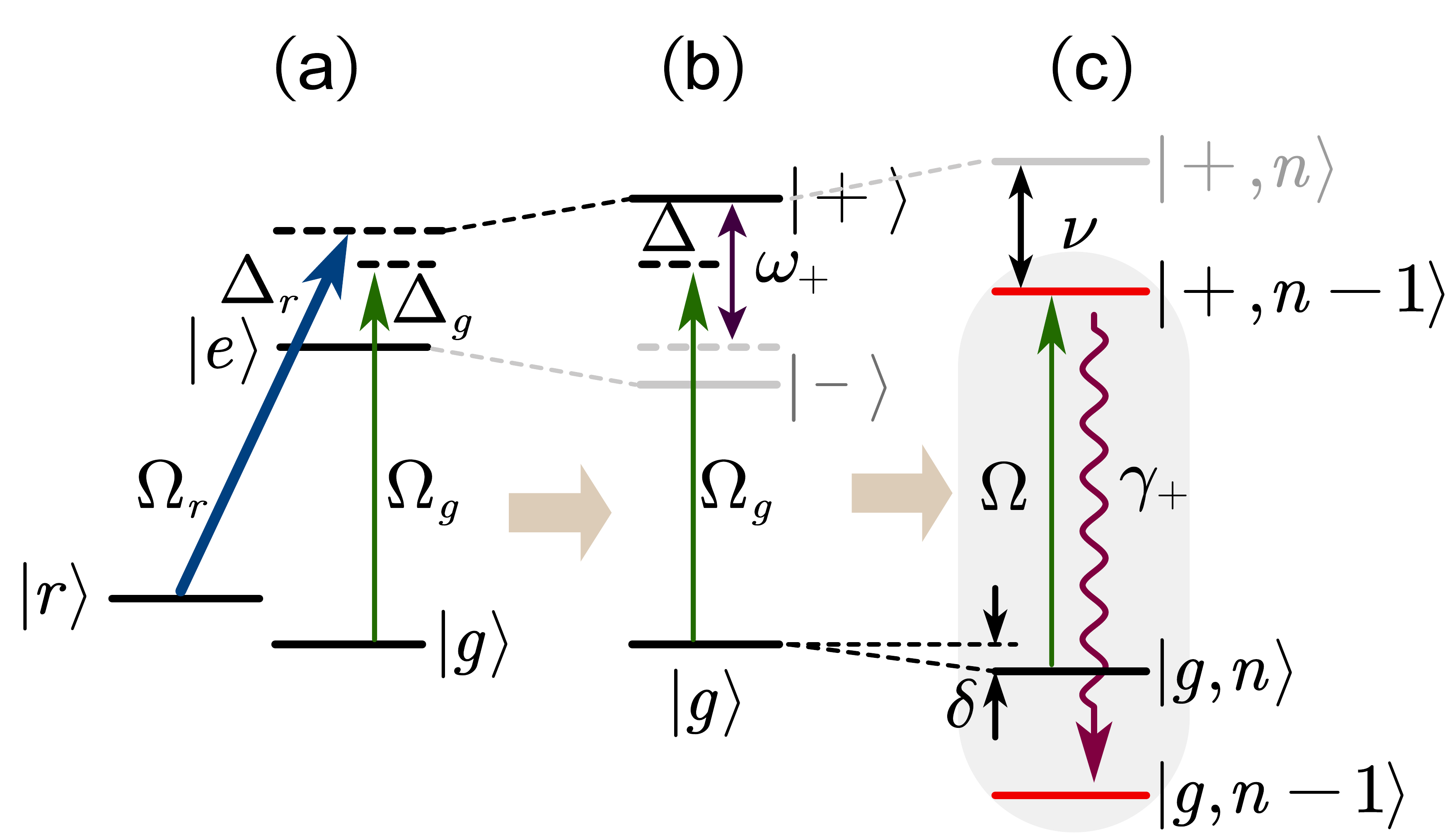}
	\caption{(a) Levels and transitions of the EIT cooling scheme (found in many species used for ion trapping). (b) The dressed levels what the cooling laser $\Omega_g$ couples. (c) When the cooling laser is near resonant with the red sideband transition of the dressed state $|+\rangle$, then this EIT cooling model can be equivalent to the model we discussed in example II.}
	\label{fig:EIT}
\end{figure}

For the EIT cooling method discussed in \cite{Double-EIT, Morigi2000, Zhang2021}, we find that the optimal parameter selection could be explained by the gap at LEP. As shown in Fig.\ref{fig:EIT} (a), the detuned laser of frequency $\omega_r$ and Rabi frequency $\Omega_r$, couples the transition $|r\rangle \leftrightarrow |e\rangle$ with detuning $\Delta_r = \omega_r - \omega_{er}$. It leads to two dressed states $|+\rangle$ and $|-\rangle$ shown in Fig.\ref{fig:EIT} (b) with energy $\omega_{+}=\Delta_r + \delta_r, \omega_{-} =  - \delta_r$ \cite{Cohen-Tannoudji2004, Morigi2000}, respectively. Here, $\delta_r=(\sqrt{\Omega_r^2 + \Delta_r^2}-|\Delta_r|)/2$ is the ac Stark shift induced by the couppling laser $\Omega_r$.

If we turn the detuning frequency $\Delta_g$ of $\Omega_g$ closing to $\Delta_r$, then $|+\rangle = \sin{\phi}|e\rangle + \cos{\phi}|r\rangle$ ($\tan \phi = \Omega_r/(\sqrt{\Omega_r^2 + \Delta_r^2} + \Delta_r) $) is chosen to replace $|e\rangle$ in the sideband cooling model (see Fig.\ref{fig:EIT} (c)). Here we just replace
\begin{eqnarray}
\gamma &\rightarrow& \gamma_+ = \gamma_e \sin^2{\phi} = \frac{\gamma_e}{2}(1-\frac{\Delta_r}{\sqrt{\Omega_r^2 + \Delta_r^2} }),\\
\Omega &\rightarrow& \eta \sin{\phi} \Omega_g, \\
\Delta &\rightarrow& \omega_+ - \Delta_g.
\end{eqnarray}
Then the optimal condition of $g_{\mathrm{max}}$ shown in Eq. (\ref{eq:condition}) will become to

\begin{eqnarray}
\frac{\eta^2}{2}(1-\frac{\Delta_r}{\sqrt{\Omega_r^2 + \Delta_r^2} }) \Omega_g^2 + (\frac{\sqrt{\Omega_r^2 + \Delta_r^2} +\Delta_r}{2}-\Delta_g)^2 = \nu^2.\nonumber\\
\end{eqnarray}

If we ignore the higher order terms $\mathcal O(\eta^2)$, Eq. (22) can be rewritten as
\begin{eqnarray}
 \frac{\sqrt{\Omega_r^2 + \Delta_r^2} +\Delta_r}{2}-\Delta_g = \delta_r + \Delta_r - \Delta_g  = \nu.
\end{eqnarray}
It is in accordance with the generalized cooling condition given in experiment works of Ref. \cite{Zhang2021, Double-EIT}, thus our LEP method offers a fresh understanding of optical cooling condition.

\section{Discussion}

In summary, we have studied how to engineer relaxation dynamics of Markovian open quantum systems with an arbitrary initial state. Our results have shown, for an arbitrary initial state, a speed-up relaxation can be achieved by setting the parameters of the system at LEP, where the slowest decay mode degenerates with a faster decay mode. In addition, our LEP-based accelerated approach can also be applied to accelerate the relaxation to stationarity in Floquet dissipative quantum dynamics. We have shown the relaxation processes can be dramatically faster than the static case by periodically modulating the dissipation strength. Finally, We have demonstrated the applications of our method for speeding up cooling processes in ground state cooling of trapped ions. In a broader view, our ideas may be still instructive for optimal parameter options to accelerate the cooling process even for simultaneous cooling of multiple phonon modes in a ion crystal. Therefor, our method would be in general and would facilitate the optical parameters setting in the experiments with open quantum many-body systems. Together with well-developed techniques of engineering quantum states, our work provides a powerful tool for exploring and utilizing true quantum LEP effects as examples of engineered relaxation dynamics \cite{Bechhoefer2021}.

\begin{acknowledgments}
This work was supported by the Natural Science Foundation of Hunan Province of China (Grants No. 2023JJ30626, 2022RC1194, 2023JJ10052), National Nature Science Foundation of China (Grants No. 12174448, 12074433, 12004430).
\end{acknowledgments}

\begin{widetext}

\begin{appendix}
\counterwithin{figure}{section}
\section{ Liouvillian spectrum of a dissipative three-level system }

We consider a simple dissipative three-level system of Fig.~\ref{fig:model_1}(b), and the dynamics is described by a Lindblad master equation
\begin{eqnarray}
\mathcal L \rho = -i [H, \rho] + J\rho J^{\dagger} - \frac{1}{2}\{J^{\dagger}J,\rho\},
\end{eqnarray}
with the Hamiltonian $H=\Omega/2 (|c\rangle \langle b|+ |b\rangle \langle c |)$ and a jump operator $J=\sqrt \gamma|a\rangle \langle c|$.

To study the Liouvilian spectra and LEPs, we first represent the Liouvillian superoperator $\mathcal L$ in a matrix form by recasting the above master equation as a matrix differential equation for the vectorized state of the density operator $\rho$. With the definitions that $|a\rangle = (1,0,0)^T, |b\rangle = (0,1,0)^T, |c\rangle = (0,0,1)^T$, the  Liouvillian superoperator is given by
\begin{eqnarray}
\mathcal L = \left(
\begin{array}{ccccccccc}
 0 & 0 & 0 & 0 & 0 & 0 & 0 & 0 & \gamma  \\
 0 & 0 & \frac{i \Omega }{2} & 0 & 0 & 0 & 0 & 0 & 0 \\
 0 & \frac{i \Omega }{2} & -\frac{\gamma }{2} & 0 & 0 & 0 & 0 & 0 & 0 \\
 0 & 0 & 0 & 0 & 0 & 0 & -\frac{i \Omega}{2}  & 0 & 0 \\
 0 & 0 & 0 & 0 & 0 & \frac{i \Omega }{2} & 0 & -\frac{i \Omega}{2}  & 0 \\
 0 & 0 & 0 & 0 & \frac{i \Omega }{2} & -\frac{\gamma }{2} & 0 & 0 & -\frac{i \Omega}{2}  \\
 0 & 0 & 0 & -\frac{i \Omega}{2} & 0 & 0 & -\frac{\gamma }{2} & 0 & 0 \\
 0 & 0 & 0 & 0 & -\frac{i \Omega}{2} & 0 & 0 & -\frac{\gamma }{2} & \frac{i \Omega }{2} \\
 0 & 0 & 0 & 0 & 0 & -\frac{i \Omega}{2}  & 0 & \frac{i \Omega }{2} & -\gamma  \\
\end{array}
\right)
\end{eqnarray}

The eigenvalues of $\mathcal L$ are
\begin{eqnarray}
\lambda_0 = 0,
\lambda_{1(2)}=-\frac{1}{4} (\gamma-\kappa),
  \lambda_{3(4)}=-\frac{1}{4} \left(\gamma+\kappa\right),
\lambda_{5,6}=-\frac{1}{2} \left(\gamma\mp\kappa\right),
\lambda_{7(8)}=-\frac{\gamma}{2}.
\end{eqnarray}
with $\kappa=\sqrt{\gamma ^2-4 \Omega ^2}$. Both the right and left eigenmatrices of the Liouvillian superoperators can be constructed to be Hermitian, they are
\begin{align}
R_0 &=\rho_{ss}= \left( \begin{array}{ccc} 1 & 0 & 0 \\ 0&0&0\\0&0&0\end{array}\right),&
R_{1} &\propto \left(
\begin{array}{ccc}
 0 & -\frac{i \left(\gamma +\kappa\right)}{2 \Omega } & 1 \\
 \frac{i \left(\gamma +\kappa\right)}{2 \Omega } & 0 & 0 \\
 1 & 0 & 0 \\
\end{array}
\right),&
R_{2} &\propto \left(
\begin{array}{ccc}
 0 & \frac{ \left(\gamma +\kappa\right)}{2 \Omega } & i \\
 \frac{\left(\gamma +\kappa\right)}{2 \Omega } & 0 & 0 \\
 -i & 0 & 0 \\
\end{array}
\right),\nonumber\\
R_{3}&\propto\left(
\begin{array}{ccc}
 0 & -\frac{i \left(\gamma -\kappa\right)}{2 \Omega } & 1 \\
 \frac{i \left(\gamma -\kappa\right)}{2 \Omega } & 0 & 0 \\
 1 & 0 & 0 \\
\end{array}
\right),&
R_{4}&\propto\left(
\begin{array}{ccc}
 0 & \frac{\left(\gamma -\kappa\right)}{2 \Omega } & i \\
 \frac{ \left(\gamma -\kappa\right)}{2 \Omega } & 0 & 0 \\
 -i & 0 & 0 \\
\end{array}
\right),&
R_5&\propto\left(
\begin{array}{ccc}
 -\frac{2 \gamma }{\gamma - \kappa } & 0 & 0 \\
 0 & \frac{2\gamma}{\gamma - \kappa}-1 & \frac{i (\gamma +\kappa )}{2 \Omega } \\
 0 & -\frac{i (\gamma +\kappa )}{2 \Omega } & 1 \\
\end{array}
\right),\nonumber\\
R_6&\propto\left(
\begin{array}{ccc}
 -\frac{2 \gamma }{\gamma +\kappa } & 0 & 0 \\
 0 & \frac{2 \gamma }{\gamma +\kappa }-1 & \frac{i (\gamma -\kappa )}{2 \Omega } \\
 0 & -\frac{i (\gamma -\kappa ) }{2 \Omega  } & 1 \\
\end{array}
\right),&
R_7&\propto\left(
\begin{array}{ccc}
 -2 & 0 & 0 \\
 0 & 1 & \frac{i \gamma }{2\Omega } \\
 0 & -\frac{i \gamma }{2\Omega } & 1 \\
\end{array}
\right),&
R_8&\propto\left(
\begin{array}{ccc}
 0 & 0 & 0 \\
 0 & 0 & 1 \\
 0 & 1 & 0 \\
\end{array}
\right),
\end{align}
\begin{align}
L_0 &= \left( \begin{array}{ccc} 1 & 0 & 0 \\ 0&1&0\\0&0&1\end{array}\right),&
L_{1} &\propto \left(
\begin{array}{ccc}
 0 & \frac{i \left(\gamma +\kappa\right)}{2 \Omega } & 1 \\
 -\frac{i \left(\gamma +\kappa\right)}{2 \Omega } & 0 & 0 \\
 1 & 0 & 0 \\
\end{array}
\right),&
L_{2} &\propto \left(
\begin{array}{ccc}
 0 & \frac{ \left(\gamma +\kappa\right)}{2 \Omega } & -i \\
 \frac{\left(\gamma +\kappa\right)}{2 \Omega } & 0 & 0 \\
 i & 0 & 0 \\
\end{array}
\right),\nonumber\\
L_{3}&\propto\left(
\begin{array}{ccc}
 0 & \frac{i \left(\gamma -\kappa\right)}{2 \Omega } & 1 \\
 -\frac{i \left(\gamma -\kappa\right)}{2 \Omega } & 0 & 0 \\
 1 & 0 & 0 \\
\end{array}
\right),&
L_{4}&\propto\left(
\begin{array}{ccc}
 0 & \frac{\left(\gamma -\kappa\right)}{2 \Omega } & -i \\
 \frac{ \left(\gamma -\kappa\right)}{2 \Omega } & 0 & 0 \\
 i & 0 & 0 \\
\end{array}
\right),&
L_5&\propto\left(
\begin{array}{ccc}
 0 & 0 & 0 \\
 0 & \frac{2\gamma}{\gamma - \kappa}-1 & -\frac{i (\gamma +\kappa )}{2 \Omega } \\
 0 & \frac{i (\gamma +\kappa )}{2 \Omega } & 1 \\
\end{array}
\right),\nonumber\\
L_6&\propto\left(
\begin{array}{ccc}
 0 & 0 & 0 \\
 0 & \frac{2 \gamma }{\gamma +\kappa }-1 & -\frac{i (\gamma -\kappa )}{2 \Omega } \\
 0 & \frac{i (\gamma -\kappa ) }{2 \Omega  } & 1 \\
\end{array}
\right),&
L_7&\propto\left(
\begin{array}{ccc}
 0 & 0 & 0 \\
 0 & 1 & -\frac{i \gamma }{2\Omega } \\
 0 & \frac{i \gamma }{2\Omega } & 1 \\
\end{array}
\right),&
L_8&\propto\left(
\begin{array}{ccc}
 0 & 0 & 0 \\
 0 & 0 & 1 \\
 0 & 1 & 0 \\
\end{array}
\right).
\end{align}
If the initial state is in this subspace ${\{|b\rangle,|c\rangle\}}$ , it is easily get that $a_{1\sim4}=\mathrm{Tr}[L_{1\sim4}\rho_{in}]=0$. It means that the coeflicients of subspace ${\{|b\rangle,|c\rangle\}}$ decomposition into $L_{1\sim4}$ are all vanished.

\section{Liouvillian spectrum of the subsystem of ground state cooling process }

For the subsystem $\{|e\rangle |1\rangle, |e\rangle |0\rangle, |g\rangle |1\rangle, |g\rangle |0\rangle\}$, we calculate the spectrum of its superoperator $\mathcal L$ and get
\begin{align}
\lambda_0 &= 0,&
\lambda_1&=\lambda_2^*=-\frac{\gamma -\kappa'}{4}+i\frac{\alpha}{2},&
\lambda_3&=\lambda_4^*=-\frac{\gamma +\kappa'}{4}+i\frac{\alpha}{2},\nonumber\\
\lambda_5&=\lambda_6^*=-\frac{\gamma }{2}+i\alpha,&
\lambda_{7(8)}&=-\frac{2\gamma \mp\epsilon}{4},&
\lambda_{9(10)}&=-\frac{2\gamma \mp\epsilon'}{4},\nonumber\\
\lambda_{11}& = \lambda_{12}^*= -\frac{3\gamma -\kappa'}{4}-i\frac{\alpha}{2},&
\lambda_{13} &= \lambda_{14}^*= -\frac{3\gamma +\kappa'}{4}-i\frac{\alpha}{2},&
\lambda_{15} &= -\gamma,
\end{align}
with
\begin{eqnarray}
\kappa'&=& \sqrt{(\gamma +2 i \beta)^2-4 \Omega^2},\\
\alpha&=& \Delta +2 \text{$\delta $}+\nu,\\
\epsilon&=& \sqrt{-2 \sqrt{\left(4 \left(\beta ^2+\Omega ^2\right)+\gamma ^2\right)^2-16 \gamma ^2 \Omega ^2}-8 \beta ^2+2 \gamma ^2-8 \Omega ^2},\\
\epsilon'&=& \sqrt{2 \sqrt{\left(4 \left(\beta ^2+\Omega ^2\right)+\gamma ^2\right)^2-16 \gamma ^2 \Omega ^2}-8 \beta ^2+2 \gamma ^2-8 \Omega ^2},
\end{eqnarray}
where $\beta = 2 \delta +\Delta -\nu$.
The corresponding eigenmatrixs are
\begin{align*}
R_0 &=\left(
\begin{array}{cccc}
 0 & 0 & 0 & 0 \\
 0 & 0 & 0 & 0 \\
 0 & 0 & 0 & 0 \\
 0 & 0 & 0 & 1 \\
\end{array}
\right),\nonumber\\
R_1 &\propto\left(
\begin{array}{cccc}
 0 & 0 & 0 & 0 \\
 0 & 0 & 0 & \frac{2 \Omega}{2 i \beta +\gamma +\kappa'^*} \\
 0 & 0 & 0 & 1 \\
 0 & \frac{2 \Omega}{-2 i \beta +\gamma +\kappa' } & 1 & 0 \\
\end{array}
\right),&
R_2&\propto\left(
\begin{array}{cccc}
 0 & 0 & 0 & 0 \\
 0 & 0 & 0 & -\frac{2 i \Omega}{2 i \beta +\gamma +\kappa'^* } \\
 0 & 0 & 0 & -i \\
 0 & \frac{2 i \Omega}{-2 i \beta +\gamma +\kappa' } & i & 0 \\
\end{array}
\right),\nonumber\\
R_3&\propto\left(
\begin{array}{cccc}
 0 & 0 & 0 & 0 \\
 0 & 0 & 0 & -\frac{2\Omega}{-2 i \beta -\gamma +\kappa'^*} \\
 0 & 0 & 0 & 1 \\
 0 & -\frac{2 \Omega}{2 i \beta -\gamma +\kappa' } & 1 & 0 \\
\end{array}
\right),&
R_4 &\propto \left(
\begin{array}{cccc}
 0 & 0 & 0 & 0 \\
 0 & 0 & 0 & \frac{2 i \Omega}{-2 i \beta -\gamma +\kappa'^*} \\
 0 & 0 & 0 & -i \\
 0 & -\frac{2 i \Omega}{2 i \beta -\gamma +\kappa' } & i & 0 \\
\end{array}
\right),\nonumber\\
R_5 &\propto \left(
\begin{array}{cccc}
 0 & 0 & 0 & 1 \\
 0 & 0 & 0 & 0 \\
 0 & 0 & 0 & 0 \\
 1 & 0 & 0 & 0 \\
\end{array}
\right),&
R_6 &\propto \left(
\begin{array}{cccc}
 0 & 0 & 0 & -i \\
 0 & 0 & 0 & 0 \\
 0 & 0 & 0 & 0 \\
 i & 0 & 0 & 0 \\
\end{array}
\right),\nonumber\\
R_7 &\propto\left(
\begin{array}{cccc}
 0 & 0 & 0 & 0 \\
 0 & -\frac{2 \gamma -\epsilon }{4 \gamma } & -\frac{\epsilon  \Omega }{\gamma  (4 i \beta +\epsilon )} & 0 \\
 0 & -\frac{\epsilon  \Omega }{\gamma  (\epsilon -4 i \beta )} & -\frac{2 \gamma +\epsilon }{4 \gamma } & 0 \\
 0 & 0 & 0 & 1 \\
\end{array}
\right),&
R_8&\propto \left(
\begin{array}{cccc}
 0 & 0 & 0 & 0 \\
 0 & -\frac{2 \gamma +\epsilon }{4 \gamma } & \frac{\epsilon  \Omega }{\gamma  (4 i \beta -\epsilon)} & 0 \\
 0 & \frac{\epsilon  \Omega }{\gamma  (-4 i \beta -\epsilon )} & -\frac{2 \gamma -\epsilon }{4 \gamma } & 0 \\
 0 & 0 & 0 & 1 \\
\end{array}
\right),\nonumber\\
R_{9}&\propto\left(
\begin{array}{cccc}
 0 & 0 & 0 & 0 \\
 0 & -\frac{2 \gamma -\epsilon'}{4 \gamma } & -\frac{\epsilon' \Omega }{\gamma  (4 i \beta +\epsilon')} & 0 \\
 0 & -\frac{\epsilon' \Omega }{\gamma  (\epsilon'-4 i \beta )} & -\frac{2 \gamma +\epsilon'}{4 \gamma } & 0 \\
 0 & 0 & 0 & 1 \\
\end{array}
\right),&
R_{10}&\propto \left(
\begin{array}{cccc}
 0 & 0 & 0 & 0 \\
 0 & -\frac{2 \gamma +\epsilon'}{4 \gamma } & \frac{ \epsilon'\Omega}{\gamma  (-\epsilon'+4 i \beta )} & 0 \\
 0 & \frac{\epsilon'\Omega}{\gamma  (-\epsilon'-4 i \beta )} & -\frac{2 \gamma -\epsilon'}{4 \gamma } & 0 \\
 0 & 0 & 0 & 1 \\
\end{array}
\right),\nonumber\\
R_{11}&\propto\left(
\begin{array}{cccc}
 0 & -1 & -\frac{-2 i \beta +\gamma +\kappa' }{2 \Omega} & 0 \\
 -1 & 0 & 0 & \frac{2 \Omega }{2 i \beta -\gamma +\kappa' } \\
 -\frac{2 i \beta +\gamma +\kappa'^*}{2 \eta  \Omega _g} & 0 & 0 & 1 \\
 0 & \frac{2 \Omega }{-2 i \beta -\gamma +\kappa'^*} & 1 & 0 \\
\end{array}
\right),\nonumber\\
R_{12}&\propto\left(
\begin{array}{cccc}
 0 & -i & -\frac{i (-2 i \beta +\gamma +\kappa' )}{2 \Omega } & 0 \\
 i & 0 & 0 & \frac{2 i \Omega }{2 i \beta -\gamma +\kappa' } \\
 \frac{i (2 i \beta +\gamma +\kappa'^*)}{2\Omega } & 0 & 0 & i \\
 0 & -\frac{2 i \Omega }{-2 i \beta -\gamma +\kappa'^*} & -i & 0 \\
\end{array}
\right),\nonumber\\
R_{13}&\propto \left(
\begin{array}{cccc}
 0 & -1 & -\frac{-2 i \beta +\gamma -\kappa' }{2 \Omega} & 0 \\
 -1 & 0 & 0 & \frac{2 \Omega }{2 i \beta -\gamma -\kappa' } \\
 -\frac{2 i \beta +\gamma -\kappa'^*}{2 \eta  \Omega _g} & 0 & 0 & 1 \\
 0 & \frac{2 \Omega }{-2 i \beta -\gamma -\kappa'^*} & 1 & 0 \\
\end{array}
\right),\nonumber\\
R_{14}&\propto\left(
\begin{array}{cccc}
 0 & -i & -\frac{i (-2 i \beta +\gamma -\kappa' )}{2 \Omega } & 0 \\
 i & 0 & 0 & \frac{2 i \Omega }{2 i \beta -\gamma -\kappa' } \\
 \frac{i (2 i \beta +\gamma -\kappa'^*)}{2\Omega } & 0 & 0 & i \\
 0 & -\frac{2 i \Omega }{-2 i \beta -\gamma -\kappa'^*} & -i & 0 \\
\end{array}
\right),\nonumber\\
R_{15} &\propto\left(
\begin{array}{cccc}
 1 & 0 & 0 & 0 \\
 0 & -1 & 0 & 0 \\
 0 & 0 & -1 & 0 \\
 0 & 0 & 0 & 1 \\
\end{array}
\right).\nonumber
\end{align*}

\begin{figure*}[tb]
\includegraphics[scale=0.7]{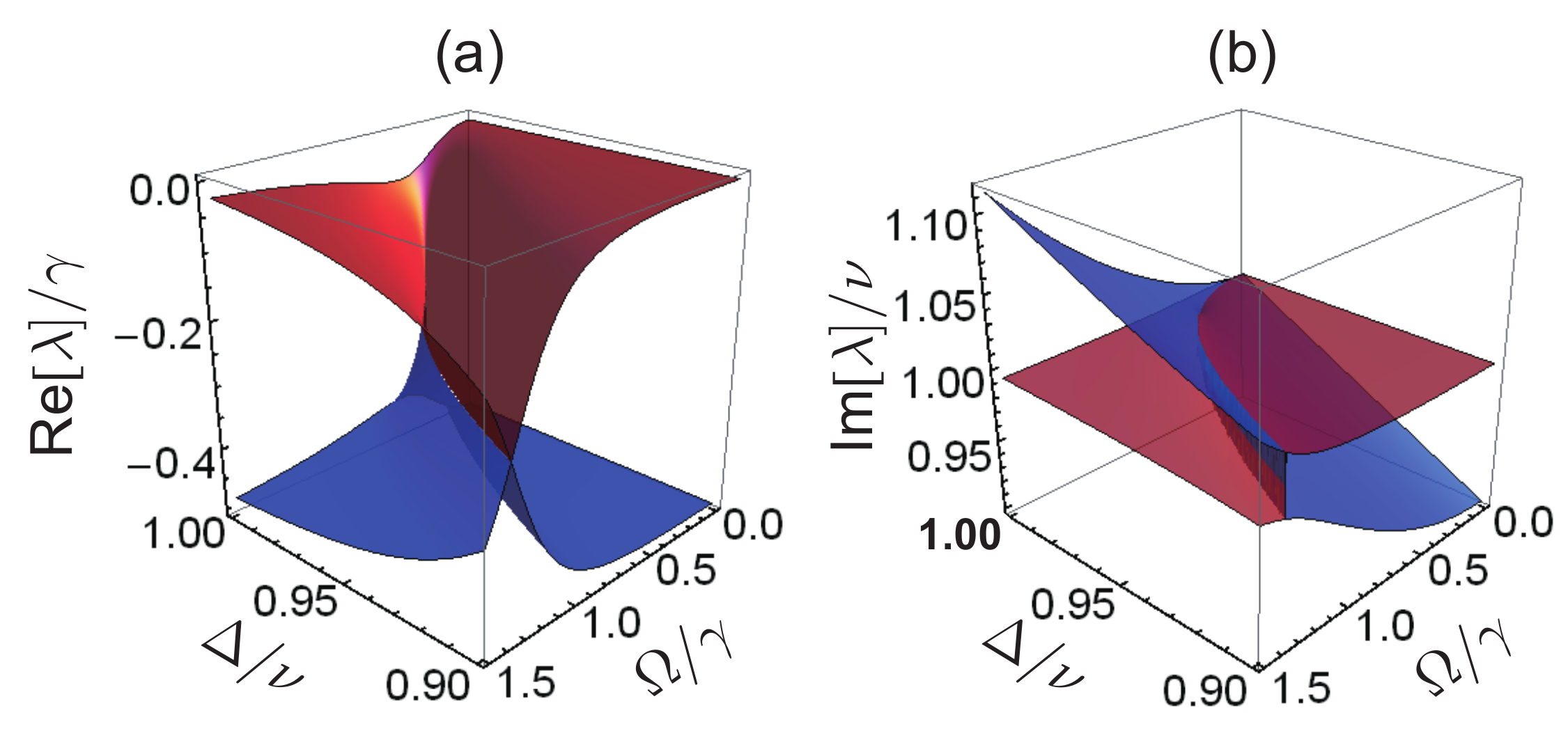}
	\caption{(a) Real and (b)imaginary parts of $\lambda_2$ (red) and $\lambda_4$ (blue).}
	\label{fig:gap3d}
\end{figure*}

We are interested with the low lying eigenvalues, especially $\lambda_1$ which determines the spectral gap $g = \frac{1}{4}(\gamma - \kappa')$. We can see that when $\beta=2 \delta +\Delta -\nu = 0 $ and $\gamma = 2 \Omega$, we get $\kappa' =0$ and $\lambda_{1(2)}=\lambda_{3(4)}$, $\lambda_{11(12)}=\lambda_{13(14)}$ (see Fig. \ref{fig:gap3d}). As a result, we can find sets of simultaneous second-order LEP.

\end{appendix}
\end{widetext}

\bibliography{bibfile}
\end{document}